\documentclass[fleqn,usenatbib]{mnras}

\usepackage{newtxtext,newtxmath}
\usepackage[T1]{fontenc}

\DeclareRobustCommand{\VAN}[3]{#2}
\let\VANthebibliography\thebibliography
\def\thebibliography{\DeclareRobustCommand{\VAN}[3]{##3}\VANthebibliography}

\usepackage{graphicx}	% Including figure files
\usepackage{amsmath}	% Advanced maths commands
\usepackage{longtable}
\usepackage {multirow}
\usepackage{threeparttable}
\usepackage{paralist}
\usepackage{booktabs}
\usepackage{xcolor}
\usepackage{ulem}

\title[Primordial Cloud and Turbulence]{Clumpy Structures within the Turbulent Primordial Cloud}

\author[Tang and Chen]{
Ching-Yao Tang,$^{1,2}$\thanks{E-mail: r08222041@g.ntu.edu.tw}
Ke-Jung Chen,$^{1}$
\\
$^{1}$Institute of Astronomy and Astrophysics, Academia Sinica, Taipei 10617, Taiwan\\
$^{2}$Department of Physics, National Taiwan University, Taipei 10617, Taiwan\\
}

\date{Accepted 2024 February 28. Received 2024 January 24; in original form 2023 June 19}

\pubyear{2024}

\begin{document}
\label{firstpage}
\pagerange{\pageref{firstpage}--\pageref{lastpage}}
\maketitle

\begin{abstract}
The primordial clouds in the mini-halos hatch the first generation stars of the universe, which play a crucial role in cosmic evolution. 
In this paper, we investigate how turbulence impacts the structure of primordial star-forming clouds. Previous cosmological simulations of the first star formation predicted a typical mass of around $\mathrm{ 100 \, M_\odot}$. This conflicts with recent observations of extremely metal-poor stars, suggesting a lower mass scale of about $\mathrm{25 \,M_\odot}$. The discrepancy may arise from unresolved turbulence in the star-forming cloud, driven by primordial gas accretion during mini-halo formation in the previous simulations. To quantitatively examine the turbulence effect on the primordial cloud formation, we employ the adaptive mesh refinement code \texttt{Enzo} to model the gas cloud with primordial composition, including artificially-driven turbulence on the cloud scale and relevant gas physics. This artificially-driven turbulence utilizes a stochastic forcing model to mimic the unresolved turbulence inside mini-halos. Our results show that the turbulence with high Mach number and compressional mode effectively fragments the cloud into several clumps, each with dense cores of $\mathrm{22.7 - 174.9 \, M_\odot}$ that undergo Jeans instability to form stars. 
Fragmentation caused by intense and compressive turbulence prevents a runaway collapse of the cloud. The self-bound clumps with smaller masses in the turbulent primordial clouds suggest a possible pathway to decrease the theoretical mass scale of the first stars, further reconciling the mass discrepancy between simulations and observations.
\end{abstract}


%%%%%%%%%%%%%%%%%%%%%%%%%%%%%%%%%%%%%%%%%%%%%%%%%%

%%%%%%%%%%%%%%%%% BODY OF PAPER %%%%%%%%%%%%%%%%%%

\section{Introduction}\label{sec:Introduction}

Based on the pioneering cosmological simulations, the first generation of stars (Population III stars in \citealt{greif2006two}, Population III.1 stars in \citealt{o2008front}; hereafter Pop~III stars) brought the early universe out of the “cosmic dark ages” by emitting first light to the dark cosmos at redshift $\mathrm{z \sim 20-30}$, roughly 50 million years after the Big Bang \citep[see][for reviews]{larson2001first,bromm2009formation,bromm2013formation,greif2015numerical,norman2018simulating,yoshida2019formation}. Furthermore, Pop~III stars synthesized first heavy elements (metals) through nuclear burning in their stellar interior; later the metals were dispersed into the intergalactic medium (IGM) through supernova explosion, and they chemically enriched the pristine gas for nurturing the subsequent star formation (SF) \citep{bromm2003first,yoshida2004era,iwamoto2005first,tominaga2007supernova,wa08b,greif2010first,fg11,chen2017chem,gen2018emp,abe21chem,chen2022impact}. Pop~III stars transform the simple early universe into an ever-increasingly complex status we observe today. Understanding the Pop~III SF system in primordial gas clouds will reveal the true nature of Pop~III stars, which is the key to understanding the genesis of our universe and life.

After the Big Bang, small matter perturbations seeded by the inflation started to grow through gravitational instabilities. Eventually, they formed into the gravitationally bound structures so-called “dark matter (DM) mini-halos” with a virial mass around $\mathrm{10^5-10^6 \, M_\odot}$ and a virial temperature of $\mathrm{\sim 10^3 \, K}$. Mini-halos amassed the primordial gas and nursed the Pop~III stars \citep{tegmark1997small,bromm2002formation,yoshida2008protostar}. The primordial gas clouds were mainly made of $\mathrm{H}$, $\mathrm{He}$, and their derivatives \citep{tegmark1997small,yoshida2003simulations,yoshida2006formation} and likely contained tiny magnetic fields of $\mathrm{<  10^{-10} \, gauss}$ \citep{wagstaff2014magnetic,jedamzik2019stringent,sanati2020constraining,mckee2020magnetic}. Because of the absence of dust and metal inside the mini-halos, molecular hydrogen became the major coolant that remains effective at a temperature below $\mathrm{10^4 \, K}$ \citep[see][chap.2]{FFLTR}, pre-determining a characteristic mass of the star-forming cloud.
For a Pop~III star-forming cloud, the mass fraction of $\mathrm{H_2}$ can reach $\sim 10^{-4}-10^{-3}$ with gas density $n \sim 10^{4} \, \rm{cm^{-3}}$; this condition allows the cloud to cool down to $\mathrm{\sim 200 \, K}$ with a corresponding Jeans mass roughly equal to $\mathrm{500 - 1000 \, M_\odot}$, and the cloud will undergo runaway collapse if its mass exceeds Jeans mass \citep{abel2002formation,yoshida2003simulations}. \par

The direct observation of Pop~III stars is far beyond the capability of our large telescopes. Thus, the chemical abundance patterns from the observation of extremely metal-poor (EMP) stars, formed right after the first stars and their supernovae, have been used to probe the typical mass of Pop~III stars via their supernova yields.
The elemental abundance of the EMP stars implies that Pop~III stars are of $\mathrm{\sim 12 - 60 \, M_\odot}$ \citep{umeda2002nucleosynthesis,umeda2005variations,joggerst2009nucleosynthetic,ishigaki2018initial, chen2017emp,chiaki2018metal}. However, the previous cosmological zoom-in simulations of Pop~III SF propose that the mass function of Pop~III stars is top-heavy and broadly distributes in $\sim \mathrm{50 - 1000 \, M_\odot}$ \citep{omukai2001formation,norman2008population,hirano2014one,hirano2015primordial,hosokawa2016formation,stacy2016building}. Other researches argue that fragmentation occurs at circumstellar dick scale and results in multiples or binaries with less massive Pop III stars \citep{stacy2010first,clark2011formation,greif2011simulations,susa2019merge,wollenberg2020formation}. Here, we focus on the fragmentation at the primordial cloud scale induced by turbulence. 
We suspect that the turbulent flow in the primordial gas cloud at the mini-halo center is a missing piece of the discrepancy between observation and simulation. Previous simulations simply treat the star-forming gas inside the halos as subsonic \citep{abel2002formation,yoshida2006formation,greif2011simulations,bromm2013formation}, and they are unable to resolve the cloud-scale turbulence inside the Pop~III star-forming cloud.
Based on \citet{tseliakhovich2010relative,greif2011delay}, supersonic flow streams persisting from the recombination epoch can create intense turbulence within mini-halos, thereby altering the physical properties of the star-forming clouds.
Turbulence offers additional pressure support, preventing the cloud from undergoing catastrophic collapse and creating multiple high-density gas regions. Ultimately, these dense regions will likely give rise to Pop~III stars with less massive stellar masses, potentially explaining the EMP stars' observed chemical abundance patterns.

Given the reasons outlined above, we believe that properly modeling subtle turbulent gas structures in the Pop~III star-forming clouds has the potential to alleviate the tension between simulations and observations.
Due to a tremendous dynamic range from the IGM down to the star-forming cloud, fully resolving the entire energy-cascading process of turbulence in any cosmological simulations is currently impossible. To investigate the impact of turbulence on the primordial cloud formation, we employ a stochastic forcing model to simulate the unresolved turbulence driven by in-flowing gas during the mini-halo formation. By manipulating parameters of supersonic turbulence, we explore the influence of various turbulence behaviors on the gas dynamics inside the mini-halos and discuss the potential subsequent SF. A similar artificial turbulence approach has been employed in Smoothed Particle Hydrodynamics (SPH) simulations of Pop~III binary stellar systems by \citet{riaz2018formation}.  

The structure of this paper is organized as follows. 
In Section \ref{sec:NumericalMethod}, we first introduce the numerical methodology for simulating the turbulent primordial could. We describe the evolution of the clouds in Section \ref{sec:TurbulentCloud}; then we present their physical properties in Section\ref{sec:ClumpyStructures}.
In Section \ref{sec:Discussions}, we discuss the applications and limits of our simulations. Finally, we conclude our findings in Section \ref{sec:Conclusion}.

%%%%%%%%%%%%%%%%%%%%%%%%%%%%%%%%%%%%%%%%%%%%%%%%%%%%%%%%%%%%%%
%%%%%%%%%%%%%%%%%%%%%%%%%%%%%%%%%%%%%%%%%%%%%%%%%%%%%%%%%%%%%%

\section{Numerical Method}\label{sec:NumericalMethod}

%%%%%%%%%%%%%%%%%%%%%%%%%%%%%

\subsection{Adaptive Mesh Refinement Code \texttt{Enzo}} \label{subsec:AdaptiveMeshRefinementCode}

We use the grid-based, adaptive mesh refinement (AMR) code \texttt{Enzo} \citep{2004astro.ph..3044O, bryan2014enzo} to model the formation of primordial gas cloud with turbulence. 
The AMR technique \citep{berger1989local,bryan1999fluids,norman1999cosmological,bryan2000hybrid} automatically increases the spatial resolution if the designated physical quantities, such as density and velocity, of the grids satisfy the refinement criteria. 
\texttt{Enzo} combines the N-body schemes \citep{hockney1988computer,couchman1991mesh} for collisionless particles and the Eulerian methods for fluid dynamics; it solves the compressible Euler equations by means of the MUSCL-based method \citep{wang2008relativistic} and the second-order Runge-Kutta scheme \citep{shu1988efficient} for time integration. 
In this work, we use the piecewise linear method \citep[PLM;][]{van1979towards,colella1985efficient} and Harten-Lax-van Leer solver \citep[HLL solver;][]{toro2013riemann} to solve Riemann problems.

To modeling the Pop~III SF cloud, we include the chemistry network of primordial gas involving nine main species: $\mathrm{H,\, H^+,\, H^-,\, H_2,\, H_2^+,}$ $\mathrm{He,\, He^+,\, He^{++},}$ and $\mathrm{e^-}$ \citep{anninos1997cosmological,abel1997modelling,abel2002formation,ripamonti2004fragmentation,turk2009formation}.
The gas cooling network considers collisional excitation and ionization, radiative recombination, free-free transition, etc., for atomic $\mathrm{H}$ and $\mathrm{He}$; moreover, it includes the $\mathrm{H_2}$ cooling due to line, formation, and collision-induced emissions.
The chemistry and cooling networks are coupled with the hydrodynamic equations self-consistently. Additionally, gas self-gravity is included by coupling the hydrodynamic equations with gravitational potential calculated from Poisson's equation.  

We utilize a stochastic forcing model developed by \citet{schmidt2009numerical}. It generates turbulence with a statistically isotropic stochastic force field, smoothly accelerating fluid on large scales. 
This external force field has been written as a stochastic differential equation in Fourier $\mathbf{k}$ space with a small spread of forcing wave numbers. The solution in $\mathbf{x}$ space is included as a source term in the momentum and energy equations of hydrodynamics.
The characteristic wave number of the force field is defined as $\mathrm{k_c = 2 \pi \alpha / l_{box}}$, where $\mathrm{l_{box}}$ is the length of the simulation box size, and $\mathrm{\alpha = 2}$ in our simulations, which means the energy injection scale is roughly half of the box size.
Our simulations of driven turbulence represent only the innermost region of the Pop~III star-forming clouds, which is encompassed by a larger turbulent region. In a realistic case, the driven scale of turbulence in our scenario is much larger than the box size. This results in $\mathrm{\alpha \ll 1}$, which introduces a broad range of uncertain length scales due to the grid size limitations of practical computations \citep{schmidt2009numerical}. In a stochastic turbulence simulation, the turbulence-driven scale should be comparable to the box size. However, setting $\mathrm{\alpha = 1}$ would lead to numerical artifacts at the boundaries of the rectangular box on the largest driven scale. Therefore, we have chosen $\mathrm{\alpha = 2}$, representing half the box size, as the largest driven scale in our simulations. Besides, the nonlinear subgrid-scale (SGS) model from \citet[]{ grete2017comparative} is used to deal with the unresolved scales of turbulence.

Stochastic driven turbulence is commonly employed in modeling turbulence within contemporary star-forming clouds enriched with metals and dust grains \citep{fed08a,fed10}. Stochastic force fields propel gas flow, elevating gas temperature through compressional heating from shock collisions. Despite the energy injection from shock heating, the presence of dust and metals allows for effective gas cooling in present star-forming clouds, maintaining an isothermal state. Therefore, simulations of these clouds often assume an isothermal equation of state (EOS) for the gas.
However, in the case of primordial gas, where metals and dust cooling are absent, the primary coolant arises from the primordial gas, as discussed in the previous section. Consequently, the assumption of an isothermal EOS is no longer valid for primordial gas. In our simulations, we adopt the ideal gas EOS coupled with primordial gas cooling to dissipate excess energy from shock heating. The local gas temperatures in the simulation region can vary from $\sim 100$ to $100,000$ Kelvin, depending on the prevailing physical conditions.

%%%%%%%%%%%%%%%%%%%%%%%
\subsection{Simulation Setup} \label{subsec:SimulationSetup}

We conduct 3D simulations using the \texttt{Enzo} code on a Cartesian coordinate grid with dimensions $x$, $y$, and $z$. The physical side length of the simulation box, denoted as $\mathrm{l_{box}}$, is set at 3 pc, comparable to the size of the central region of primordial clouds. This choice aligns with the scale suggested in the previous cosmological simulations that recommend a scale of 5 pc \citep{yoshida2008protostar,bromm2009formation}.
Two cloud masses $M_{gas}$ are used in our simulations: $\rm{3397 \, M_\odot}$ and $\rm{6041 \, M_\odot}$.
The box is initially filled with uniform primordial gas (76\% H and 24\% He by mass) of densities $\rm{8.4 \times 10^{-21} \, g \, cm^{-3}}$ or $\rm{1.5 \times 10^{-20} \, g \, cm^{-3}}$, and a uniform temperature 1000 K is set based on the previous cosmological simulations of Pop~III SF \citep{greif2011simulations, hirano2015primordial}.
The root grid has $\mathrm{256^3}$ cells with up to two levels of factor-two refinement ($2^2$). The refinement criteria are based on the gas overdensity and Jeans length; the grid will be refined if gas density $\mathrm{ > 10^{-17} \, g \, cm^{-3}}$, or the number of the cells covered one Jeans length is less than 16. 
The finest grids have a spatial resolution of $\mathrm{\sim 604 \, AU}$, which is roughly the size of proto-stellar envelope ($\mathrm{\gtrsim 300 \, AU}$); therefore, the physical processes of dense core formation ($\rm{n \sim 10^{8} \, cm^{-3}}$) can be well resolved in our simulation.

%%%%%%%%%%%%%%%%%%%%%%%
\subsubsection{Physical Scenario of Modelling the Turbulence in Primordial Gas}\label{subsub:PhysicalScenario}

In our scenario, the primordial gas is accreted onto the halo center through the halo gravity during the assembly of DM mini-halos, and this process leads to gravito-turbulence. The turbulence persists until the turbulent primordial cloud becomes virialized.
Then the cloud stops amassing the primordial gas, and the sub-halo turbulence diminishes.
Meanwhile, $\rm{H_2}$ cooling dissipates the thermal energy injected from turbulence so that gas self-gravity becomes dominant.
After that, the high-density gas clusters shaped by the turbulence can grow into gravitationally bound structures that host the Pop~III SF.
To realize the above scenario, we divide our simulation into three different phases, which will be described in the following sections.

%%%%%%%%%%%%%%%%%%%%%%%%%
\subsubsection{Phase I: Stochastic Turbulence Development}\label{subsub:StochasticTurbulenceDevelopment}

Resolving the turbulence of Pop~III star-forming region from galactic scales down to protostellar objects is still beyond the envelope of modern cosmological simulations. 
Since we are interested in the inner region of primordial clouds of several pc, we adopt the stochastic method with periodic boundaries on all sides of the simulation box to model the innermost region of the mini-halo instead of evolving the turbulence cascade from the scale larger than halo. We define this early simulation stage as \textbf{Phase I}.
The original algorithm of stochastic turbulence in \texttt{Enzo} only considers the isothermal gas. We have modified this algorithm to make it applicable to the ideal gas with the primordial chemistry and cooling network. 
The stochastic force field induces gas motion through incorporating a source term in the momentum equation of hydrodynamics. Simultaneously, gas self-gravity modifies the mass distribution by updating the momentum source term with the solution of Poisson's equation. When both self-gravity and driven turbulence are considered concurrently in the simulation, they can interact, often leading to numerical instability and run crashes eventually. To overcome this problem, we deactivate the self-gravity while the stochastic force field is operating. In \textbf{Phase I} of the simulation, we assume that gas flow is primarily dominated by turbulent motion, and its self-gravity is considered negligible. 

In this phase, stochastic turbulence stirs up the primordial gas in an isotropic manner, and the associated gas chemistry and cooling coevolve with the development of turbulence.
The stochastic force field will shape the gas structure inside the primordial cloud.
To explore the impact of turbulence on the primordial star-forming cloud, we select different combinations of two turbulence parameters in accordance with \citet{schmidt2009numerical} definition.
The first turbulence parameter is characteristic Mach number $\mathcal{M}$, where $\mathcal{M}= V/c_0$, $V$ is the characteristic velocity, and $c_0$ is the initial sound speed.
The second turbulence parameter is $\mathcal{C}$ which represents the ratio between the compressional and solenoidal components in the force field. The turbulent flow becomes more compressive as $\mathcal{C}$ increases.

The previous studies of Pop~III SF suggest that the gas flow of scale $\sim 10$ pc in the mini-halo is either subsonic or transonic of $\mathcal{M} \leq 1$ \citep{abel2002formation,greif2011simulations}, resulting in a relatively minor effect of turbulence. 
However, these simulations face challenges in accurately capturing turbulent flows within the mini-halo. This limitation stems from the highly zoomed-in nature of the simulations and the constrained spatial resolution at the halo scale. Additionally, both theoretical models and observations of SF in the local and distant universe \citep{low04,lad05,kru05,tac20} highlight the significance of supersonic turbulence in influencing the mass distribution of stars. Therefore, we use the moderate supersonic flow with a $\mathcal{M} = 1-10$ on the cloud and clump scales to examine the impact of supersonic turbulence on the primordial star-forming region. We simulate this turbulence using $\mathcal{M}$ values of 2, 4, and 8 to analyze its effects.
Given the lack of data on the Pop~III star-forming sites, we draw on present-day studies \citep{2017A&A...599A..99O} indicating that the solenoidal fraction of gas flow in such regions could be as low as 0.25. Thus, we consider our turbulence models to have $\mathcal{C}$ with values 2, 3, and 4 to sample the possible Pop~III SF scenario. All of our models and their associate parameters are summarized in Table \ref{tab:ModelParameters}.

\begin{table*}
    \centering 
	
    \begin{tabular}{ c|c|c|cccc } 
    \toprule
    Model & 
    $\mathcal{M}$ &
    $\mathcal{C}$ &
    $M_{gas} \ \mathrm{[M_\odot]}$ &
    $\rho_{i} \ \mathrm{[g \, cm^{-3}]}$ &
    $M_{h} \ \mathrm{[M_\odot]}$ &
    $r_{h} \ \mathrm{[pc]}$ \\ 
    
    \midrule
    $M2C2S$ &
    \multirow{6}*{2} &
    \multirow{2}*{2} &
    \multirow{1}*{3397} &
    \multirow{1}*{$\mathrm{8.4 \times 10^{-21}}$} &
    \multirow{1}*{$\mathrm{3\times10^5}$} &
    \multirow{1}*{110}
    \\
    $M2C2D$ & ~ & 
    ~ &
    6041 &
    $\mathrm{1.5 \times 10^{-20}}$ &
    $\mathrm{3.1\times10^5}$ &
    97 
    \\ 
    $M2C3S$ & ~ &
    \multirow{2}*{3} & 
    3397 & 
    $\mathrm{8.4 \times 10^{-21}}$ &
    $\mathrm{3\times10^5}$ &
    100
    \\ 
    $M2C3D$ & ~ & 
    ~ &
    6041 &
    $\mathrm{1.5 \times 10^{-20}}$ &
    $\mathrm{3.1\times10^5}$ &
    97 
    \\
    $M2C4S$ & ~ &
    \multirow{2}*{4} & 
    3397 &  
    $\mathrm{8.4 \times 10^{-21}}$ &
    $\mathrm{3\times10^5}$ &
    100 
    \\
    $M2C4D$ & ~ &
    ~ & 
    6041 &
    $\mathrm{1.5 \times 10^{-20}}$ &
    $\mathrm{3.1\times10^5}$ &
    97
    \\

    \midrule
    $M4C2S$ &
    \multirow{6}*{4} &
    \multirow{2}*{2} &
    \multirow{1}*{3397} &
    \multirow{1}*{$\mathrm{8.4 \times 10^{-21}}$} &
    \multirow{1}*{$\mathrm{3\times10^5}$} &
    \multirow{1}*{110}
    \\
    $M4C2D$ & ~ & 
    ~ &
    6041 &
    $\mathrm{1.5 \times 10^{-20}}$ &
    $\mathrm{3.1\times10^5}$ &
    97 
    \\ 
    $M4C3S$ & ~ &
    \multirow{2}*{3} & 
    3397 & 
    $\mathrm{8.4 \times 10^{-21}}$ &
    $\mathrm{3\times10^5}$ &
    100
    \\ 
    $M4C3D$ & ~ & 
    ~ &
    6041 &
    $\mathrm{1.5 \times 10^{-20}}$ &
    $\mathrm{3.1\times10^5}$ &
    97 
    \\
    $M4C4S$ & ~ &
    \multirow{2}*{4} & 
    3397 & 
    $\mathrm{8.4 \times 10^{-21}}$ &
    $\mathrm{3\times10^5}$ &
    100 
    \\
    $M4C4D$ & ~ &
    ~ & 
    6041 &
    $\mathrm{1.5 \times 10^{-20}}$ &
    $\mathrm{3.1\times10^5}$ &
    97
    \\

    \midrule
    $M8C2S$ &
    \multirow{6}*{8} &
    \multirow{2}*{2} &
    \multirow{1}*{3397} &
    \multirow{1}*{$\mathrm{8.4 \times 10^{-21}}$} &
    \multirow{1}*{$\mathrm{3\times10^5}$} &
    \multirow{1}*{110}
    \\
    $M8C2D$ & ~ & 
    ~ &
    6041 &
    $\mathrm{1.5 \times 10^{-20}}$ &
    $\mathrm{3.1\times10^5}$ &
    97 
    \\ 
    $M8C3S$ & ~ &
    \multirow{2}*{3} & 
    3397 & 
    $\mathrm{8.4 \times 10^{-21}}$ &
    $\mathrm{3\times10^5}$ &
    100
    \\ 
    $M8C3D$ & ~ & 
    ~ &
    6041 &
    $\mathrm{1.5 \times 10^{-20}}$ &
    $\mathrm{3.1\times10^5}$ &
    97 
    \\
    $M8C4S$ & ~ &
    \multirow{2}*{4} & 
    3397 & 
    $\mathrm{8.4 \times 10^{-21}}$ &
    $\mathrm{3\times10^5}$ &
    100 
    \\
    $M8C4D$ & ~ &
    ~ & 
    6041 &
    $\mathrm{1.5 \times 10^{-20}}$ &
    $\mathrm{3.1\times10^5}$ &
    97
    \\

    \bottomrule
    \end{tabular}
    
    \caption{Model parameters. From left to right: model name, characteristic Mach number $\mathcal{M}$, ratio between the compressional and solenoidal components $\mathcal{C}$, total gas mass $M_{gas}$ in the simulation box, initial gas density $\rho_i$, corresponding virial mass $M_h$ and virial radius $r_h$ of the mini-halo in accordance with \citet{greif2011simulations,greif2012formation}.} \label{tab:ModelParameters}
    
\end{table*}

%%%%%%%%%%%%%%%%%%%%%%%%%%%%%
\subsubsection{Phase II: Turbulence Diminishing}\label{subsub:TurbulenceDiminishing}

Stochastic turbulence transitions into dynamic equilibrium as the overall gas temperature reaches its minimum due to cooling. Following this, we posit that turbulence initiates decay owing to the cessation of in-flowing gas from the halo scale.
Therefore, at this moment, we halve the characteristic Mach number every turnover time of the largest eddy $\mathrm{\tau_{eddy} \sim l_{eddy}/ v_{rms}}$, where $\mathrm{l_{eddy}}$ and $\mathrm{v_{rms}}$ are the largest eddy size (0.5 $\mathrm{l_{box}}$) and the root mean square velocity of the gas. We repeat this reduction procedure until $\mathcal{M}$ drops to 1. 
Meanwhile, a fixed potential of DM mini-halo is induced; the corresponding virial mass and radius are determined based on cosmological simulations from \citet{greif2011simulations,greif2012formation}.
The gravitational potential well of mini-halos follows the NFW profile \citep{navarro1996structure, navarro1997universal},
\begin{equation}
    \rho_{\mathrm{DM}}(r) = \frac{M_{h}}{4 \pi \big( \frac{r_{h}}{c} \big) ^3 \big[ ln(1+c)-\frac{c}{1+c} \big] } \frac{1}{\frac{r}{r_s} \big( 1+\frac{r}{r_s} \big) ^2} \, ,
\end{equation}
where $M_{h}$ and $r_{h}$ are the virial mass and virial radius of the halo, $r_s$ is the scale radius, and $c = r_{h}/r_s$ is the concentration parameter, which is 20 in this study. 
The minimum potential is positioned at the center of the simulation box, where the halo gravity attracts turbulent gas toward the halo center. The process of turbulence reduction, influenced by halo gravity as described above, is referred to as \textbf{Phase II}.

%%%%%%%%%%%%%%%%%%%%%%%%%%%%%
\subsubsection{Phase III: Dense Core Formation }\label{subsub:DenseCoreFormation}

We reduce the intensity of the driven turbulence until the dense gas reaches a state of slow collapse, determined by the virialization parameter as defined in \citep{wise2008resolving},
\begin{equation}
    \beta = \frac{3(\gamma-1)E_{th} + 2E_{k}}{E_s - U}-1,
\end{equation}
where $E_{th}$, $E_k$, $E_s$, and $U$ denote the thermal energy, kinetic energy, surface pressure work, and gravitational potential energy, respectively. 
$\gamma$ is the adiabatic index, which we adopt $5/3$ for the ideal monatomic gas in the $\beta$ calculation.
Self-gravity gradually emerges as a dominant force in the system as turbulence subsides in the latter stages of \textbf{Phase II}. Hence, the gravitational binding energy within the gas $U_{b}$ and dark matter potential energy $U_{DM}$ acting on the gas are encompassed in the term $U$ to assess the instantaneous virialization status of the cloud.
When $\beta$ diminishes to zero, we deactivate the stochastic force field and activate the self-gravity of the gas, which becomes the predominant force in the weakened turbulence, fostering the growth of clumpy structures within the cloud. Additionally, the boundary condition is altered from periodic to outflow to simulate the collapse of the cloud. This final stage of the simulation can be referred to as \textbf{Phase III}.

\begin{figure*}
\centering
\includegraphics[width=1.0\linewidth]{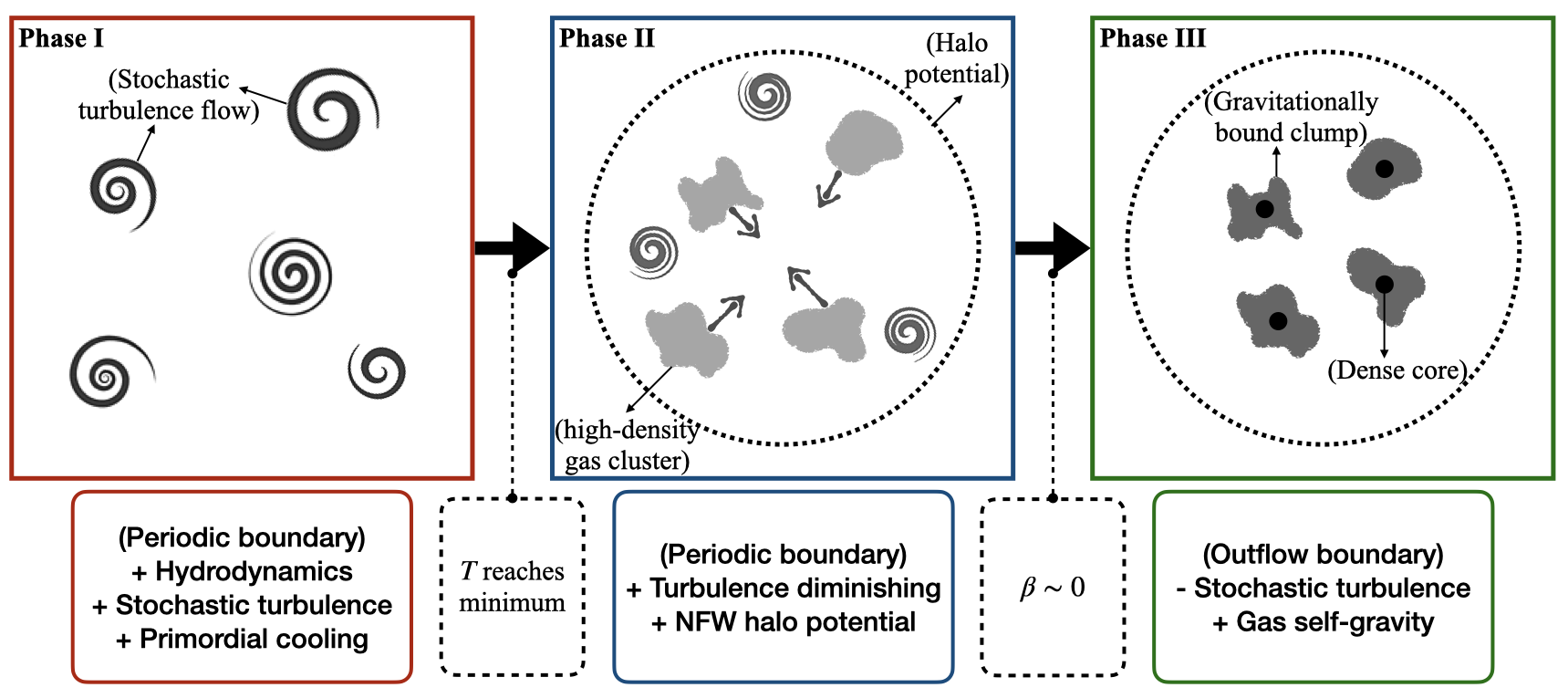} 
\caption{Flowchart of simulating the turbulent primordial cloud. We separate the simulation into three phases, and the physical processes included or excluded in each phase are shown in the text box below the cartoon. The transition point between two consecutive phases is according to the criterion stated in the dashed-line text box.}
\label{FlowChart}
\end{figure*}

During \textbf{Phase III}, the high-density gas clusters grow in mass and become gravitationally bound.
If the maximum gas density in the simulation exceeds $\mathrm{10^{-15} \, g \, cm^{-3}}$, the three-body reaction will rapidly convert most of the hydrogen atom into molecular form \citep{turk2010effects}.
While fully molecular gas has the capacity to cool and condense rapidly, we encounter limitations in evolving our simulations further, primarily stemming from spatial resolution constraints and the absence of small-scale microphysics.
Thus, we have to terminate the simulations when the maximum density reaches $\mathrm{\sim 10^{-16} \, g \, cm^{-3}}$. Finally, we summarize the simulation procedure, including three major phases in Figure \ref{FlowChart}.  

%%%%%%%%%%%%%%%%%%%%%%%%%%%%%%%%%%%%%%%%%%%%%%%%%%%%%%%%%%%%%%
%%%%%%%%%%%%%%%%%%%%%%%%%%%%%%%%%%%%%%%%%%%%%%%%%%%%%%%%%%%%%%
\section{Evolution of the Turbulent Primordial Clouds} \label{sec:TurbulentCloud}

%%%%%%%%%%%%%%%%%%%%%%%%%%%%%
\subsection{Chemical and Thermal Evolution}\label{subsec:ChemicalThermalEvolution}

\begin{figure*}
\begin{minipage}[t]{0.33\textwidth}
\centering
\includegraphics[width=1.0\linewidth]{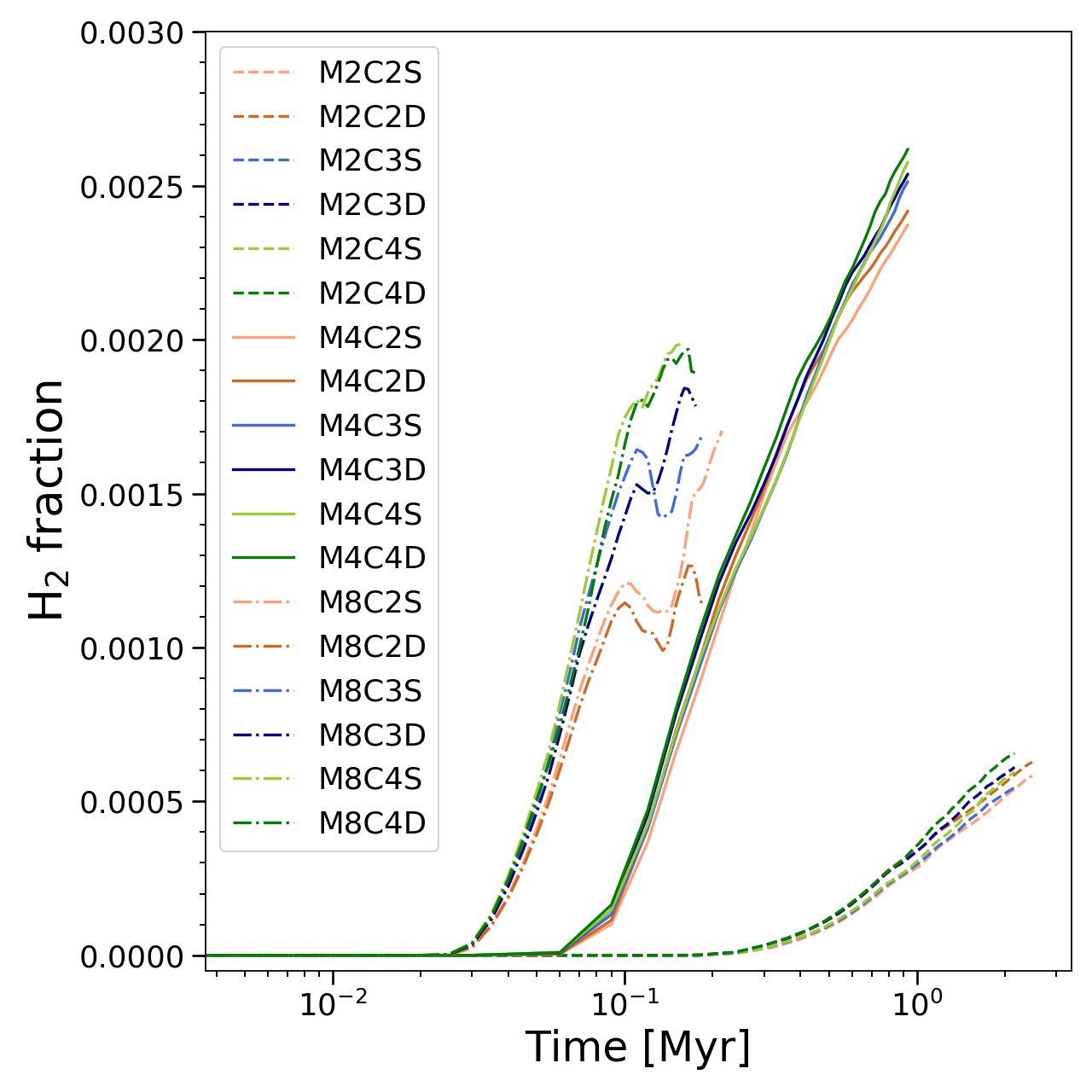}
\end{minipage}
\begin{minipage}[t]{0.33\textwidth}
\centering
\includegraphics[width=1.0\linewidth]{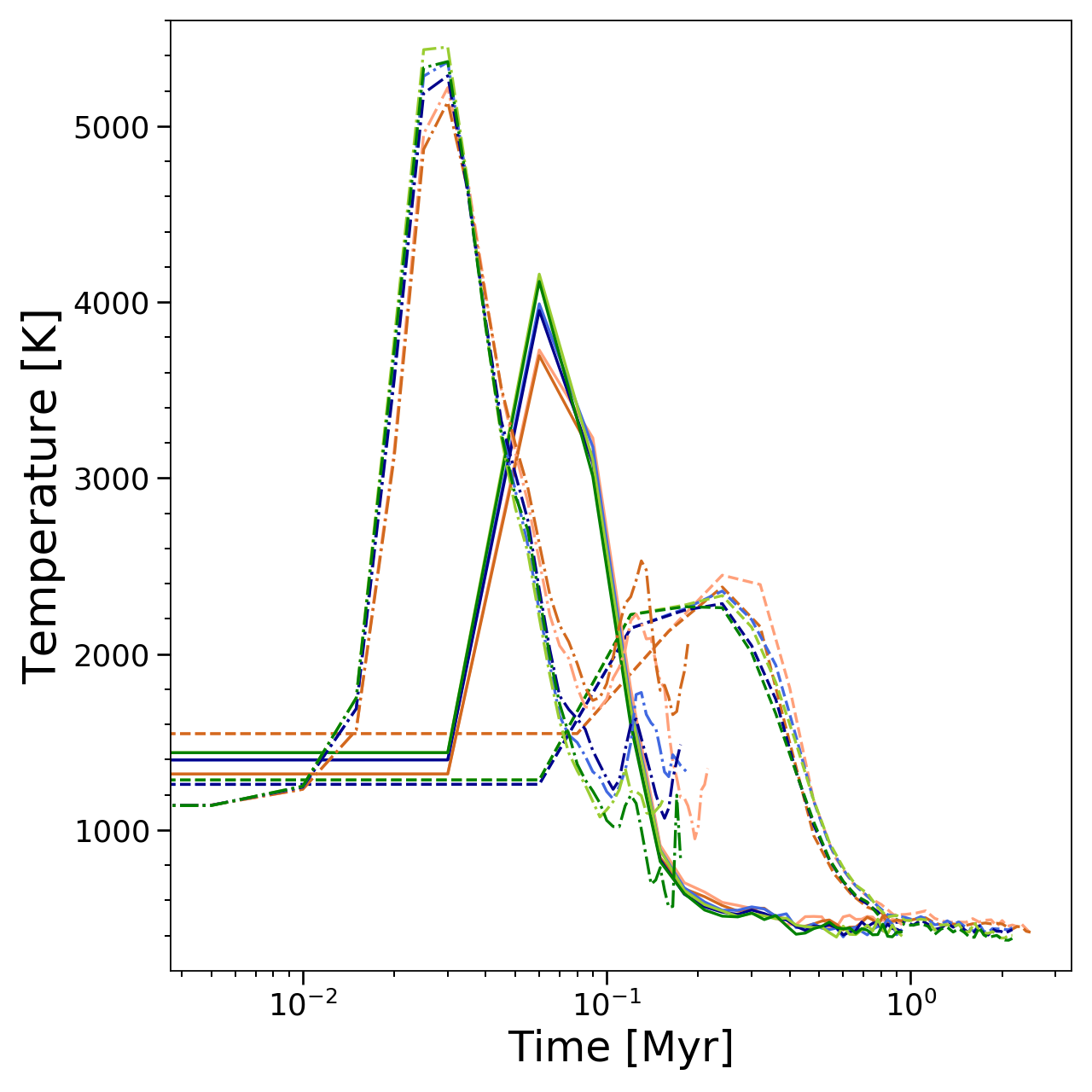}
\end{minipage}
\begin{minipage}[t]{0.33\textwidth}
\centering
\includegraphics[width=1.0\linewidth]{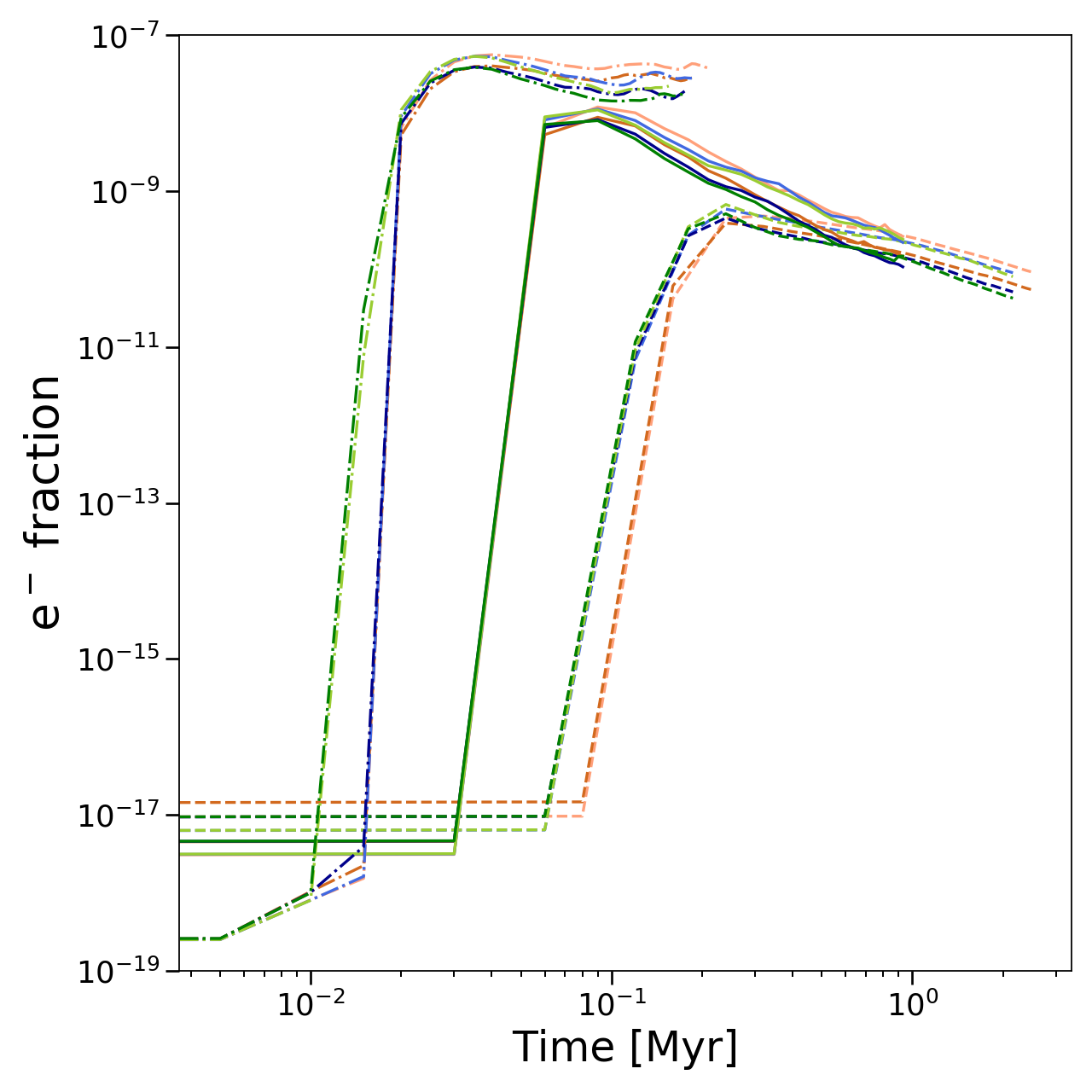}
\end{minipage}
\caption{Evolution of overall mass-weighted $\mathrm{H_2}$ mass fraction (left), temperature (middle), and electron mass fraction (right) during the \textbf{Phase I} simulation. The evolutionary tracks of $\rm{H_2}$ fraction, temperature, and electron fraction group mostly with $\mathcal{M}$, and they are less sensitive to  $\mathcal{C}$. Some large fluctuations occur in $\rm{H_2}$ fraction and temperature of $\mathcal{M} = 8$ models are due to the collision dissociation.}

\label{TimeEvol}
\end{figure*}

In the left panel of Figure \ref{TimeEvol}, we illustrate the temporal progression of the $\mathrm{H_2}$ mass fraction (hereafter referred to as $\mathrm{H_2}$ fraction) during \textbf{Phase I}.
As shown in the figure, $\mathrm{H_2}$ fraction rapidly increases from zero to $\sim 10^{-4}$ within the first $0.03 - 0.5$ million years after the simulation begins; then, the molecular hydrogen cooling effect becomes substantial.
The models sharing the same Mach number $\mathcal{M}$ show similar growth trajectories in terms of $\mathrm{H_2}$ fraction. Generally, higher $\mathcal{M}$ models tend to produce more $\mathrm{H_2}$ fraction, as they generate highly compressed flows and create denser gas regions that favor the formation of $\mathrm{H_2}$.
In the fixed $\mathcal{M}$ models with varying $\mathcal{C}$, the deviation among evolutionary tracks becomes more significant as $\mathcal{M}$ increases. Our findings suggest that $\mathrm{H_2}$ formation is more sensitive to $\mathcal{M}$ than to $\mathcal{C}$. However, unlike the consistently growing trajectories observed in $\mathcal{M}=2$ and 4, those in $\mathcal{M}=8$ exhibit fluctuations at later times due to the collisional dissociation of $\mathrm{H_2}$ by free electrons \citep{abel1997modelling},
\begin{equation}
H_2 + e^- \rightarrow
H + H + e^-.
\end{equation}

While stochastic turbulence evolves, driven eddies inject kinetic energy into the primordial clouds. Subsequently, a portion of gas kinetic energy transforms into thermal energy through fluid compression, resulting in increased gas density, temperature, and $\mathrm{H_2}$ fraction. As illustrated in the middle panel of Figure \ref{TimeEvol}, temperatures rise from about 1000 to $\rm{2200 - 5400 \, K}$ within the initial $0.03 - 0.3$ million years after the simulations commence. Following the peak temperatures, they start to decline due to enhanced $\mathrm{H_2}$ cooling, eventually reaching equilibrium ($\mathrm{\lesssim \, 400 \, K}$ in $\mathcal{M}=2 \, \mathrm{and} \, 4$ models). Like the evolutionary tracks of $\mathrm{H_2}$ fraction, temperature tracks in fixed $\mathcal{M}$ models exhibit similarity. However, temperatures in $\mathcal{M}=8$ models display considerable variation at later times, settling around $\mathrm{\sim \, 1000 - 2000 \, K}$—higher than the equilibrium temperatures of other $\mathcal{M}$ models. This discrepancy is attributed to the strong turbulence in $\mathcal{M}=8$ models continuously injecting more energy into the cloud than it can dissipate. Despite the temperature fluctuations in $\mathcal{M}=8$ models, we select a snapshot at approximately $\rm{1000 \, K}$ as the starting point for their \textbf{Phase II} simulation.

Furthermore, we present the electron mass fraction (hereafter referred to as electron fraction) of the gas in the right panel of Figure \ref{TimeEvol}. Similar to the evolutionary tracks of $\mathrm{H_2}$ fraction and temperature, the electron fraction experiences a rapid increase from roughly $10^{-18}$, reaching $10^{-10} - 10^{-8}$, and the timing of this increase aligns with that of $\mathrm{H_2}$ fraction and temperature. This abrupt rise in the electron fraction is attributed to collisional ionization induced by compressional heating from shocks. For $\mathcal{M}=8$ models, the electron fraction undergoes a minor decrease after reaching its peak because of the relatively high background temperature.

\begin{figure*}
\centering
\includegraphics[width=0.83\linewidth]{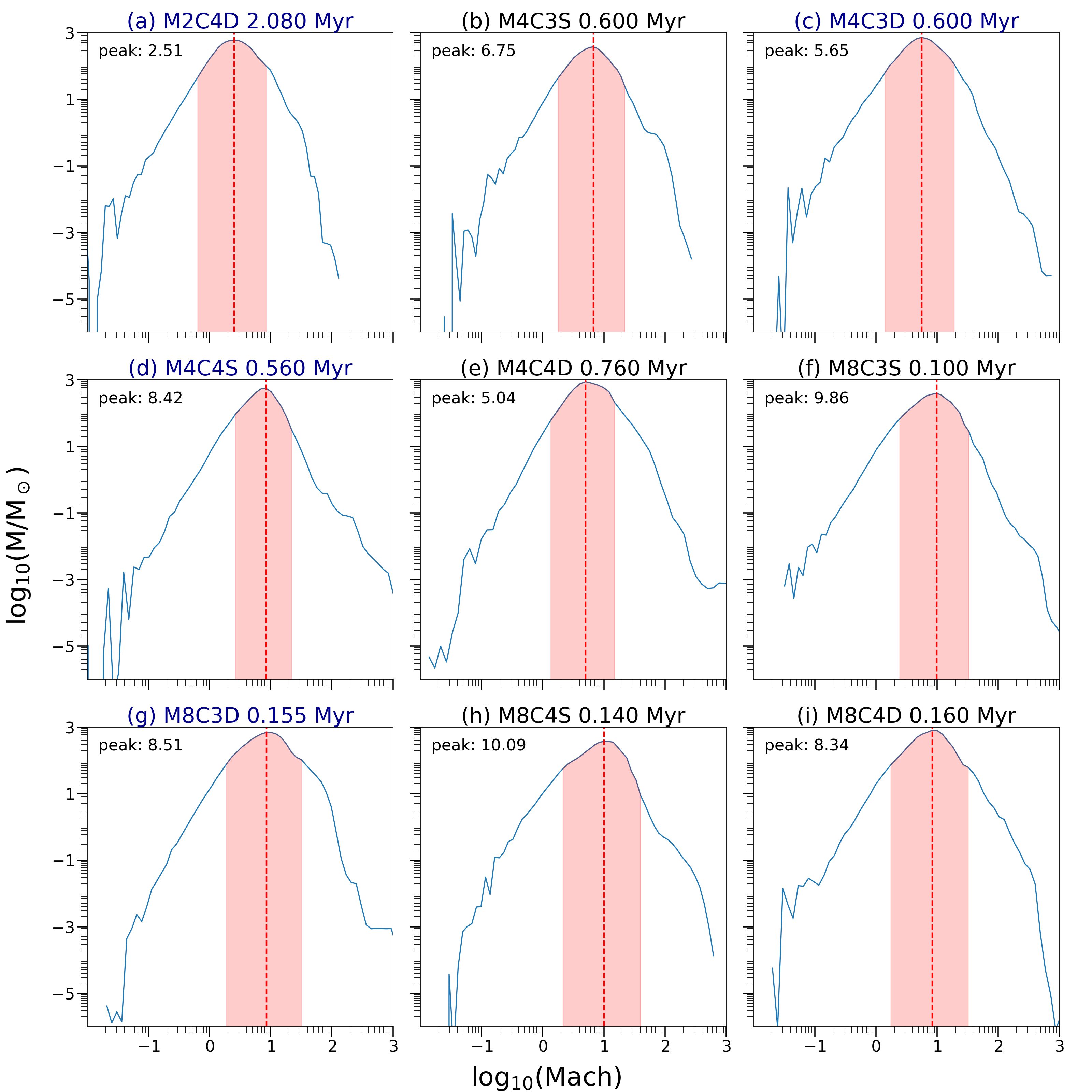} 
\caption{Gas mass distribution as a function of Mach number at the end of the \textbf{Phase I} simulation. The location of the peak indicated by the red-dashed line shows the Mach number bin containing the most mass, and its value is marked on the upper left corner in each panel. Besides, the red-shaded area covers 95\% of the gas mass around the peak.
At this moment, the peak Mach number roughly corresponds to $\mathcal{M}$ of the model.
The models with panel titles in dark blue color do not form gravitationally bound structure in \textbf{Phase III}.}
\label{mass-mach-histogram1}
\end{figure*}

In Figure \ref{mass-mach-histogram1}, the gas mass distribution is depicted as a function of Mach number at the end of the \textbf{Phase I} simulation. Given that our turbulence system is non-isothermal, the gas temperature within the simulation domain spans from 10 to $\mathrm{10^5}$ K, resulting in the Mach numbers of gas flow ranging from 0.1 (subsonic) to several hundreds (hypersonic). However, the Mach numbers of over $95\%$ of the gas in each model closely align with the assigned characteristic $\mathcal{M}$. Gas with extremely high and low Mach numbers collectively constitutes less than $1\%$ of the total mass.

%%%%%%%%%%%%%%%%%%%%%%%%%%%%%
\subsection{Virialization of the Clouds}\label{subsec:Virialization}

\begin{figure*}
\centering
\includegraphics[width=0.83\linewidth]{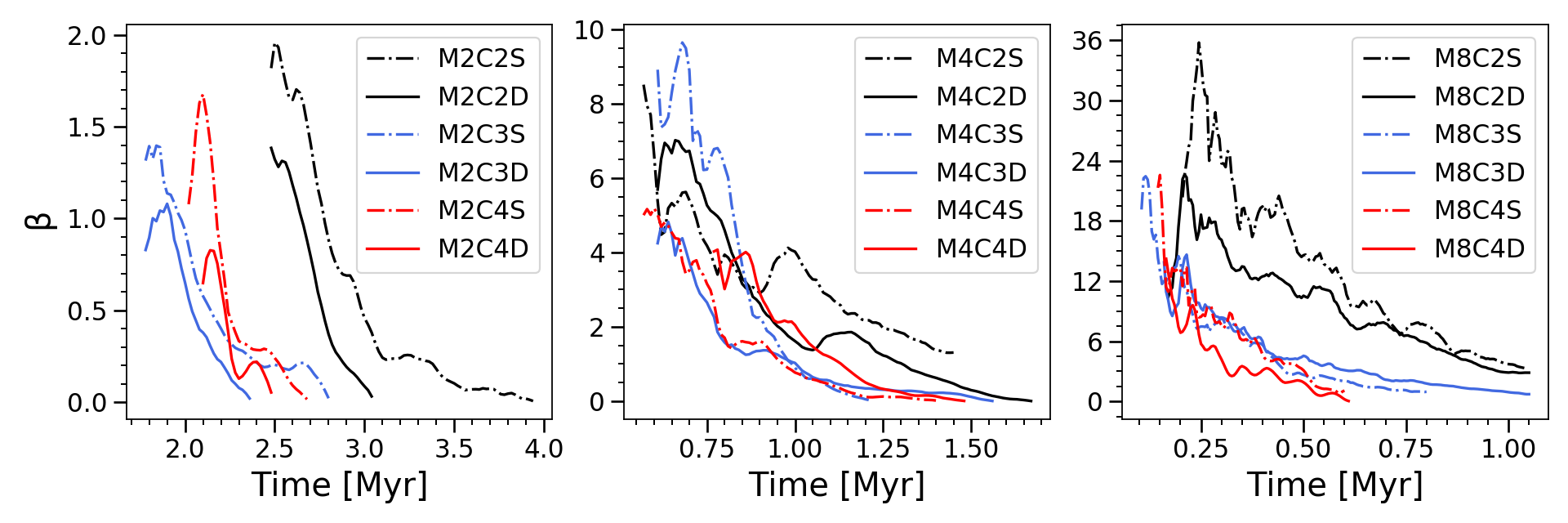} 
\caption{Evolution of virialization parameter $\beta$ during the \textbf{Phase II} simulation. Three panels correspond to $\mathcal{M} =$ 2, 4, and 8, respectively. The decline in $\beta$ over time is due to the gradual reduction in turbulence strength,  
the $\rm{H_2}$ cooling, inclusion of a DM potential. At the end of \textbf{Phase II}, $\beta$ reaches approximately zero, except for the three $\mathcal{C} = 2$ models.}
\label{beta-time}
\end{figure*}

\begin{figure*}
\centering
\includegraphics[width=1.0\linewidth]{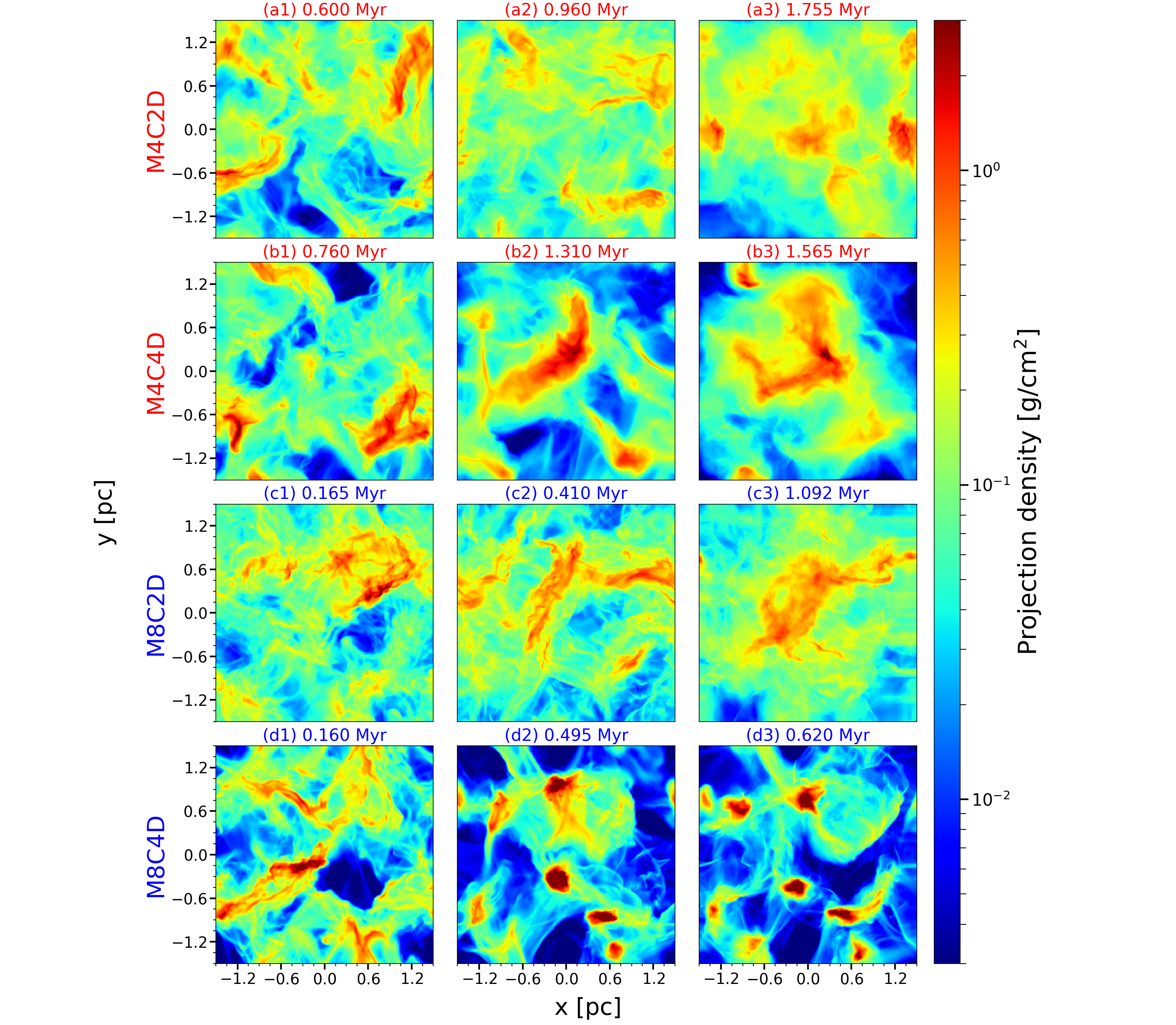} 
\caption{Evolution of gas projection density (column density) in the simulation. From left to right columns show the snapshots of gas column densities at \textbf{Phase I}, \textbf{Phase II}, and \textbf{Phase III}. 
Red and blue labels represent $\mathcal{M}=4$ and 8, correspondingly.
The gas configurations evolve from a relatively isotropic distribution to a more concentrated structure throughout the simulation. In the case of $\mathcal{C} = 4$ models, high-density gas clusters are formed in the third phase. Additionally, model $M8C4D$ forms multiple clusters due to turbulence fragmentation.}
\label{pro-dens-time}
\end{figure*}

We present the evolution of the virialization parameter during the \textbf{Phase II} simulation in Figure \ref{beta-time}.
At the outset of this phase, models with higher $\mathcal{M}$ values exhibit larger $\beta$ due to a stronger turbulence left from \textbf{Phase I}. However, $\beta$ subsequently decreases rapidly, influenced by the following effects:
\begin{enumerate}
    \item Reducing the strength of the stochastic forcing field.
    \item Inclusion of the DM halo potential.
    \item Effective cooling of $\rm{H_2}$ in the dense region.
\end{enumerate}
Upon including the DM halo potential and reducing the strength of the stochastic forcing field gradually, $\beta$ values decline to zero, except for the models with high $\mathcal{M}$ but low $\mathcal{C}$. Higher $\mathcal{M}$ clouds exhibit stronger turbulence energy ($E_k$ and $E_{th}$) and resist global collapse. Furthermore, less compressive turbulence in the lower $\mathcal{C}$ models leads to a relatively sparse gas configuration, resulting in insufficient $\rm{H_2}$ cooling compared to the energy injection. Consequently, $\beta$ values in these models fail to drop below 0.

To visualize the evolution of gas structure, we present spatial distributions of the projected gas density (column density) in Figure \ref{pro-dens-time}. At the end of the \textbf{Phase I} simulation (column 1 in the figure), turbulent flow develops into a relatively isotropic configuration due to the nature of the stochastic forcing field. Subsequently, during \textbf{Phase II} (column 2), after the decay of turbulence and under the gravitational influence of the DM halo potential, the gas flow starts concentrating around the central region, forming clumpy structures within the turbulent clouds. Eventually, in \textbf{Phase III}, as shown in the third column, some compact regions become Jeans unstable and collapse further into high-density objects.

%%%%%%%%%%%%%%%%%%%%%%%%%%%%%%%%%%%%%%%%%%%%%%%%%%%%%%%%%%%%%%
%%%%%%%%%%%%%%%%%%%%%%%%%%%%%%%%%%%%%%%%%%%%%%%%%%%%%%%%%%%%%%
\section{Clumpy Structures in the Turbulence}\label{sec:ClumpyStructures}

In the following sections, we examine the physical parameters of the turbulent Pop~III primordial clouds to investigate the criteria for forming the dense clumpy structures, the candidates of the Pop~III star-forming sites.
We then delve into the physical properties of the gravitationally bound clumps formed in the simulations, analyzing their chemi-thermal composition and developmental history. Lastly, we define dense cores originating from collapsing clumps and discuss the potential outcomes of SF within them.

\begin{figure*} 
\centering
\includegraphics[width=0.82\linewidth]{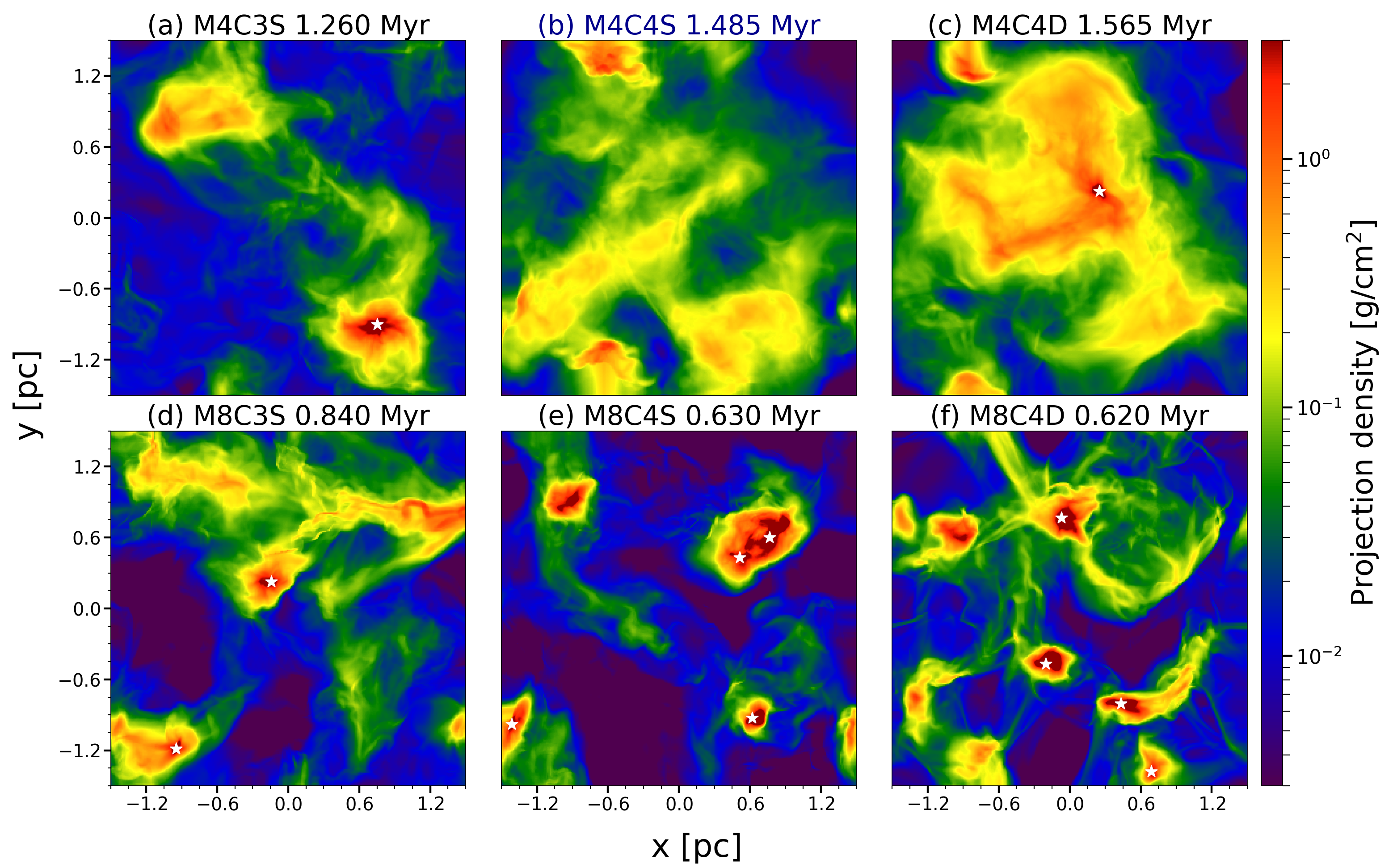} 
\caption{The density projection snapshots include all models that have formed Gravitationally Bound Clusters (GBC) at the end of the entire simulation. The white star marks indicate local density maxima with significant in-falling gas surrounding them, considered as cores (discussed in Section \ref{subsec:Clumps'Properties} and \ref{subsec:DenseCores}). In the case of $\mathcal{M} = 4$, only a single clump has formed in the simulation. On the other hand, $\mathcal{M} = 8$ models form 2 to 4 clumps. Plot (b) of $M4C4S$ fails to form any clump at the end.}
\label{proj-clump-final}
\end{figure*}

%%%%%%%%%%%%%%%%%%%%%%%%%%%%%
\subsection{Turbulence Effects}\label{subsec:TurbulenceEffect}

To assess the impact of turbulence properties on gas structure, we commence by comparing the projection density plots among different models in Figure \ref{pro-dens-time}. In \textbf{Phase I} (column 1 in the figure), the gas configurations appear similar; however, clouds with higher $\mathcal{C}$ exhibit greater structural complexity, evident in the presence of more relatively low-density areas (depicted in blue regions in the plots).
After the diminishment of turbulence (column 2), only the dense regions in $\mathcal{C} = 4$ clouds show an increase in density (rows b and d). In contrast, in $\mathcal{C} = 2$ models, previously high-density objects become more diffusive as turbulence strength decreases. Notably, several local gas clusters with relatively high density form exclusively in models with high $\mathcal{M}$ and $\mathcal{C}$, such as $M8C4D$.
In \textbf{Phase III} (column 3), the gas self-gravity emerges as the dominant force shaping the evolution of gas clusters, solidifying these high-density structures, particularly in $\mathcal{C} = 4$. Nevertheless, forming high-density gas clusters is challenging in $\mathcal{C} = 2$ models (rows a and c).

We present the density projections for the models that have formed gravitationally bound clumps (GBCs or clumps in short) at the end of the entire simulation in Figure \ref{proj-clump-final}. Each GBC consists of at least one dense core (indicated by a star mark in the figure), which will be discussed in Section \ref{subsec:DenseCores}. Among the $\mathcal{M}=4$ models, two of them form a single clump in the gas cloud, whereas three $\mathcal{M}=8$ models create $2-4$ clumps. Additionally, extensive fragmentary structure is observed in high $\mathcal{M}$ cases. A unique clump in $M8C4S$ contains two dense cores, suggesting that clump fragmentation can occur in strong turbulence ($\mathcal{M} \geq 8$).

\begin{figure*}
\begin{minipage}[t]{0.5\textwidth}
\centering
\includegraphics[width=0.95\linewidth]{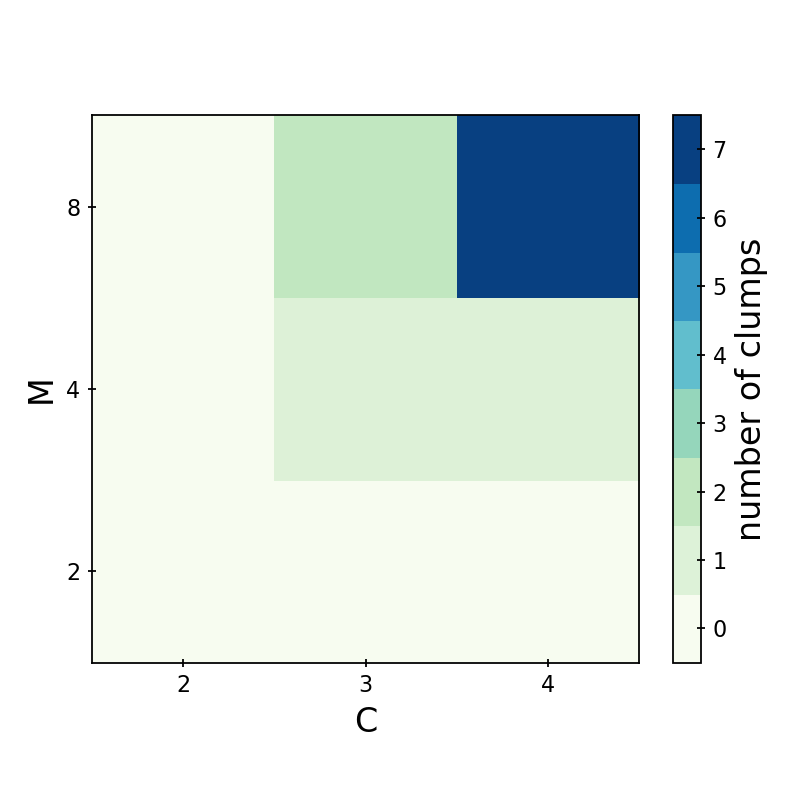}
\end{minipage}
\begin{minipage}[t]{0.45\textwidth}
\centering
\includegraphics[width=0.95\linewidth]{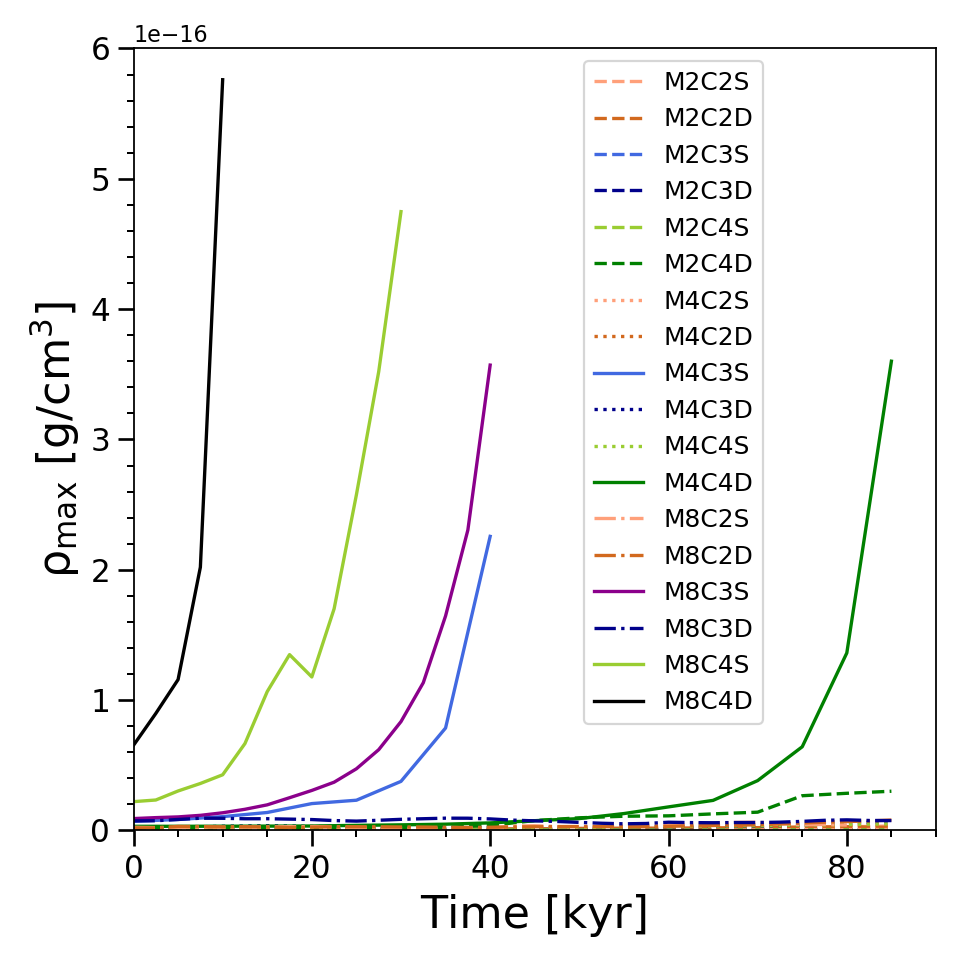}
\end{minipage}
\caption{Left panel: The correlation among clump number, $\mathcal{M}$, and $\mathcal{C}$.
Only the models with $\mathcal{M} \geq 4$ and $\mathcal{C} \geq 3$ form clumps. For $\mathcal{M} = 8$, clump number increases as $\mathcal{C}$.
Right panel: Evolution of maximum gas density in the simulation during \textbf{Phase III}. The evolution time here starts from the beginning of \textbf{Phase III}. For the models that form GBCs, their maximum densities grow rapidly above the threshold we set for the simulation termination.
}
\label{Summary}
\end{figure*}

The correlation between the number of GBCs and the turbulence properties is summarized in the left panel of Figure \ref{Summary}. Our findings indicate a positive correlation between the number of GBCs and turbulence parameters $\mathcal{M}$ and $\mathcal{C}$. Furthermore, no GBC forms in weak ($\mathcal{M} \leq 2$) or low-compression ($\mathcal{C} \leq 2$) turbulence. In other words, strong and highly compressive turbulence is likely to reshape a single cloud into clumps with smaller masses.

\begin{figure*} 
\centering
\includegraphics[width=0.82\linewidth]{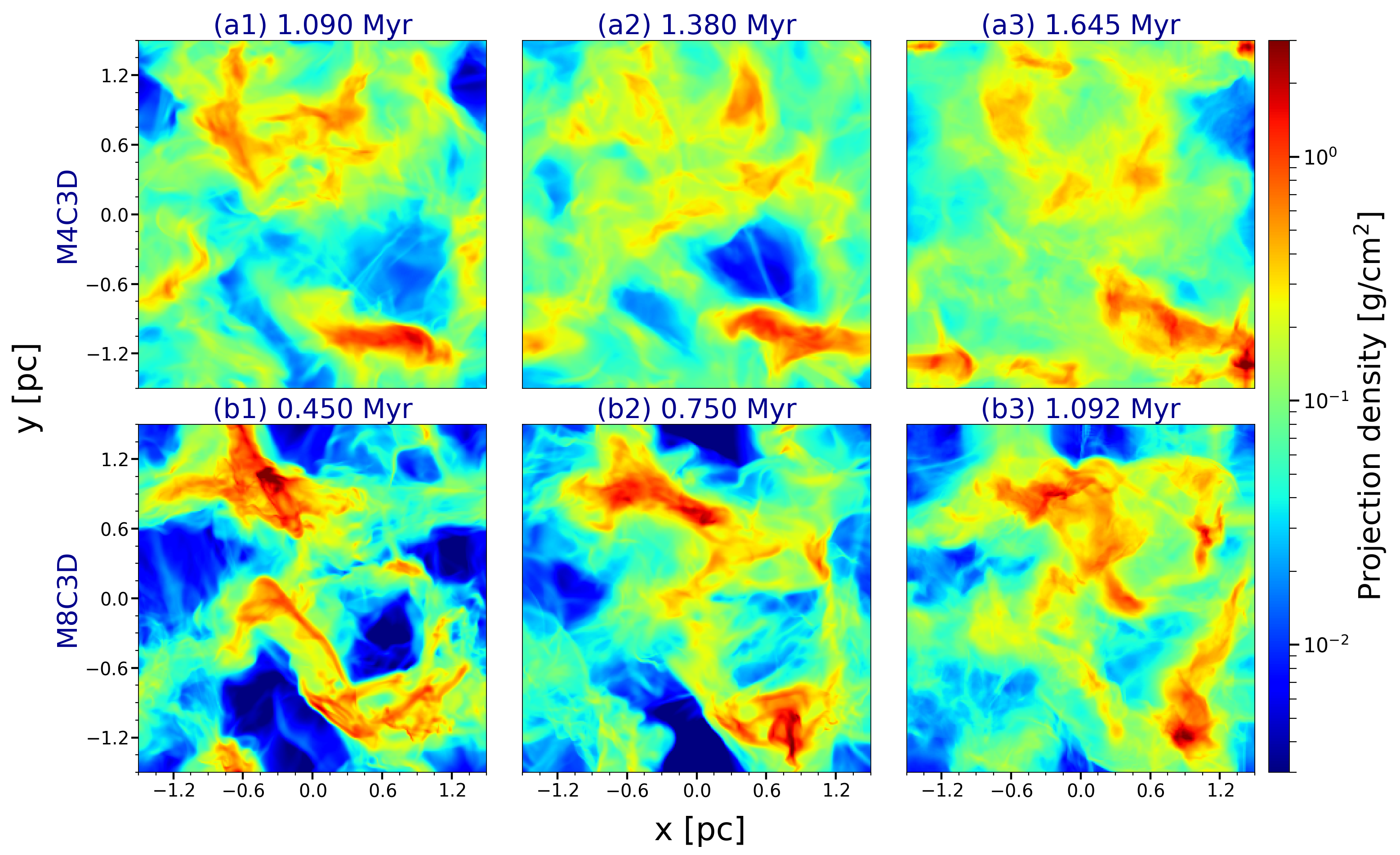} 
\caption{Density projection evolution of $M4C3D$ and $M8C3D$. Three columns from left to right correspond to three time steps: The first column shows the snapshot after two turnover time when $\mathcal{M}$ reduces to 1. The second and third columns show the snapshots of the end of \textbf{Phase II} and \textbf{Phase III} simulation. Although some high-density gas clusters form in \textbf{Phase II}, they still fail to grow into Jeans unstable objects.}
\label{pro-dens-no-clump-time}
\end{figure*}

The evolution of the maximum gas density in the \textbf{Phase III} simulations is illustrated in the right panel of Figure \ref{Summary}. Generally, only models with $\mathcal{M} \geq 4$ and $\mathcal{C} \geq 3$ meet the density threshold and form gravitationally bound clumps (GBCs). However, high $\mathcal{M}$ and $\mathcal{C}$ are necessary but insufficient conditions for GBC formation.
Once gas self-gravity becomes the dominant force in this phase, maximum densities in the models with GBCs increase rapidly due to the collapse of the densest structures within the GBCs. The densest regions in $M8C4S$ and $M8C4D$ have become Jeans unstable and undergone collapse rapidly. Turbulence in these models compresses part of the cloud into an over-dense state predominantly governed by self-gravity. For $M8C3S$, $M4C4D$, and $M4C3S$, clumps continue to grow in mass following the collapse of their dense cores.

As mentioned earlier, $\mathcal{M} \geq 4$ and $\mathcal{C} \geq 3$ are necessary but insufficient conditions for forming gravitationally bound clumps (GBCs). Our findings suggest that only models with strong and highly compressive turbulence ($\mathcal{M} = 8$ and $\mathcal{C} = 4$) can give rise to clumps in both mass scales ($M8C4S$ and $M8C4D$). On the other hand, three models—$M8C3D$, $M4C4S$, and $M4C3D$—fail to form GBCs, while their low or high mass counterpart models ($M8C3S$, $M4C4D$, and $M4C3S$) do. Therefore, GBC formation appears to be insensitive to the overall cloud mass.

In Figure \ref{pro-dens-no-clump-time}, we follow the evolution of $M8C3D$ and $M4C3D$. No significant gas accretion occurs in \textbf{Phase II} (panels a1 and a2) in $M4C3D$. Although some gas assembles at the boundaries of the simulation box after self-gravity is activated, these gas clusters are not massive enough to become Jeans unstable and undergo possible SF. In model $M8C3D$, high-density gas clusters formed early are later disrupted by stochastic turbulence (upper left corner of panel b1). Despite forming high-density gas clusters in the lower right corner of panel b2, they fail to evolve into gravitationally bound structures by the end. In summary, turbulence has both positive and negative impacts on GBC formation due to its random nature; in weaker and less compressive turbulence ($\mathcal{M} \leq 4$ and $\mathcal{C} \leq 3$), the disruptive effect could be stronger than the compressive effect, inhibiting GBC formation.

\begin{figure*} 
\centering
\includegraphics[width=0.9\linewidth]{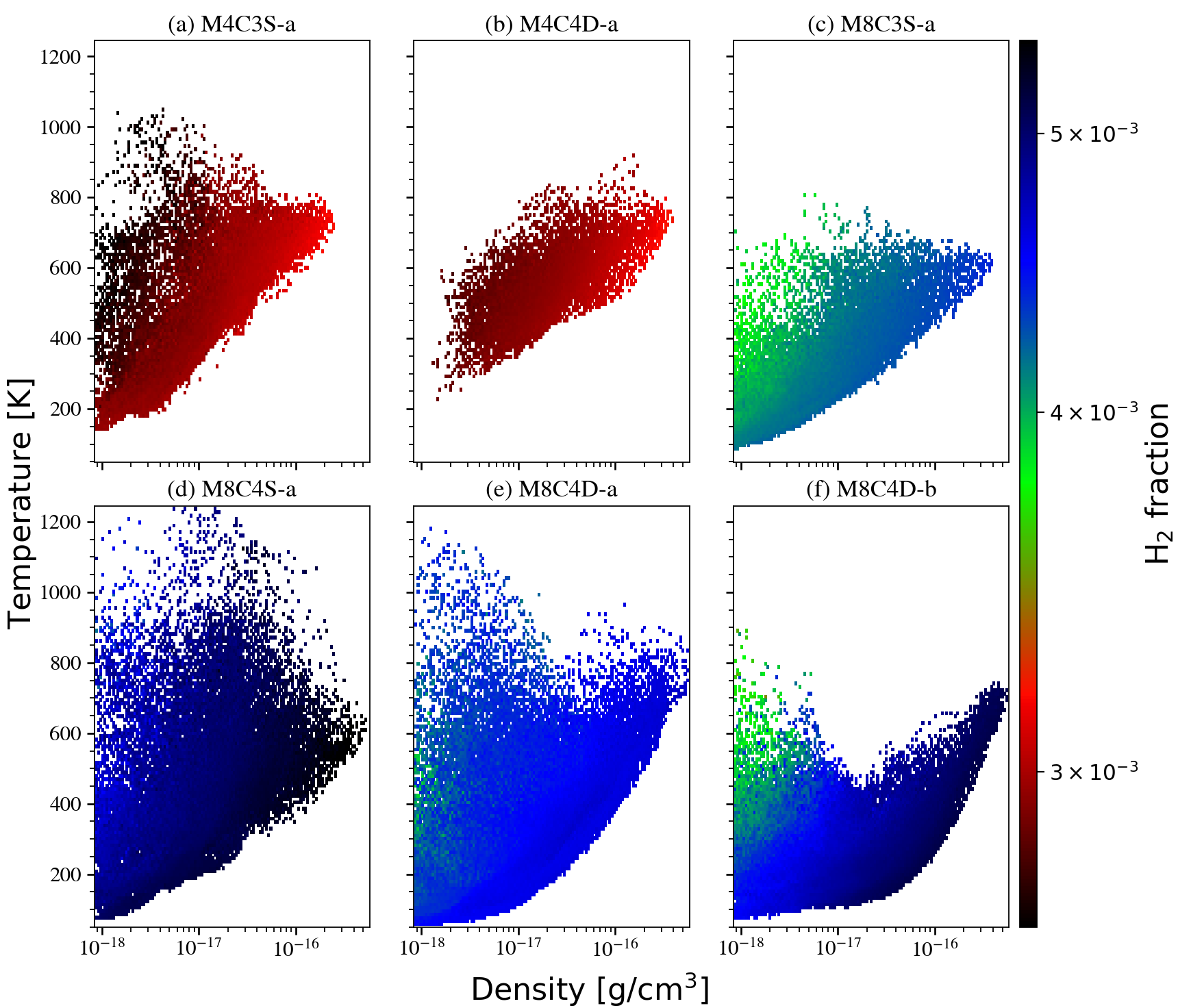} 
\caption{The density-temperature diagram with the associated $\mathrm{H_2}$ fraction of clumps at the end of the \textbf{Phase III} simulation is shown. Each panel represents a GBC, and the name and properties of each GBC can be found in Table \ref{tab:DenseCoreProperties}. In each clump, a similar $\mathrm{H_2}$ fraction is distributed in a diagonal trend from the lower-left to the upper-right. The high $\mathrm{H_2}$ fraction is distributed around higher density but lower temperature.}
\label{phase-plot}
\end{figure*}

%%%%%%%%%%%%%%%%%%%%%%%%%%%%%
\subsection{Clumps' Properties}\label{subsec:Clumps'Properties}

The temperature-density-$\mathrm{H_2}$ phase diagram of the clumps is presented in Figure \ref{phase-plot}. In general, the distribution of similar molecular hydrogen fractions in the diagram follows a diagonal trend from the lower left to the upper right. Since the cooling rate is proportional to $\mathrm{H_2}$ fraction, mass contraction leads to an increase in temperature and density under the same cooling efficiency. The fact that gas with a higher $\mathrm{H_2}$ fraction is concentrated in the lower right corner implies that denser gas can cool down to lower temperatures due to more efficient cooling.
Clumps in stronger and more compressive turbulence exhibit a higher overall $\mathrm{H_2}$ fraction, consistent with the results for the entire gas system discussed in Section \ref{subsec:ChemicalThermalEvolution}. Moreover, clumps in $\mathcal{M} = 8$ models can cool to temperatures below 100 K. For simplicity, we neglect $\mathrm{HD}$ cooling, which is the key coolant for primordial gas below 100 K in this study. In primordial gas with temperatures below 100 K, $\mathrm{HD}$ cooling surpasses $\mathrm{H_2}$ cooling and has the potential to cool the gas down to several tens of Kelvin, fostering an environment conducive to the formation of more compact structures.

The six clumps depicted in Figure \ref{phase-plot} represent the densest objects in their respective simulations. The variation in molecular hydrogen fraction is relatively small, even with a twofold difference in $\mathcal{M}$. 
The bottleneck in molecular hydrogen fraction growth suggests that turbulence has reached its limitation in catalyzing $\mathrm{H_2}$ formation. Gas self-gravity becomes essential to condense the clump further and trigger the rapid three-body reaction of hydrogen. However, such high-density regions are beyond the capability of our simulation, prompting us to terminate the simulation when the maximum density of clumps exceeds approximately $\mathrm{10^{-16} \, g/cm^{3}}$.

\begin{figure*}
\centering
\includegraphics[width=1.0\linewidth]{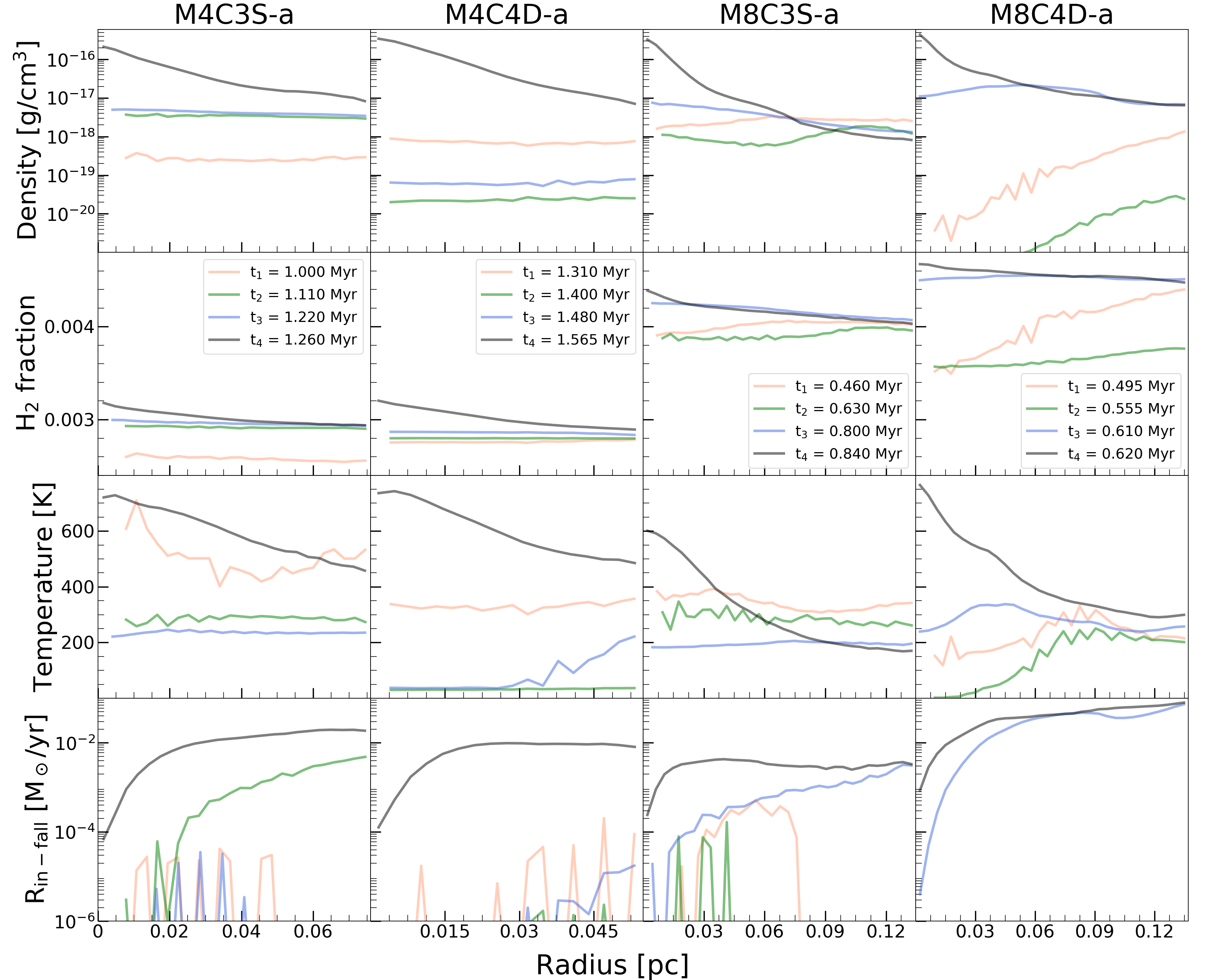} 
\caption{The figure displays the spherical-averaged profiles of density, $\mathrm{H_2}$ mass fraction, temperature, and in-fall rate for the clumps during \textbf{Phase II} and \textbf{Phase III}. 
We define four profile times as following: $\mathrm{t_1}$ is the snapshot after two turnover time when $\mathcal{M}$ reduces to 1, $\mathrm{t_2=(t_1+t_3)/2}$, $\mathrm{t_3}$ is the end of the \textbf{Phase II} simulation, and $\mathrm{t_4}$ is the end of the \textbf{Phase III} simulation.
The label of each clump corresponds to the information in Table \ref{tab:DenseCoreProperties}. The profiles illustrate that the $\rm{H_2}$ fraction increases only slightly over time. In the final density and temperature profiles, the centers of the clumps evolve into hot and dense structures, where the in-fall rate drops significantly at $r \lesssim 0.025 , \rm{pc}$.}
\label{radius-profile}
\end{figure*}

To explore the physical properties of GBCs, we approximate the clumps as spheres centered at their maximum density and calculate their 1D radial profiles of density, $\mathrm{H_2}$ fraction, temperature, and infall rate at different evolution times in Figure \ref{radius-profile}. At the final snapshot ($\mathrm{t_4}$, black lines), density and chemi-thermal configurations are similar among the inner regions of these clumps; all of them possess a relatively hot and dense structure with a little increment in $\mathrm{H_2}$ fraction.
We also observe a flat density profile of cores at the end of \textbf{Phase II} ($\mathrm{t_3}$, blue lines) in Figure \ref{radius-profile}. This profile results from shock compression and insufficient cooling at the core scale. Since self-gravity is not activated at this point, there is no other contraction force to condense the gas further, and the flat profile persists without the influence of gravity. The initial flat profile arises from the assumption that turbulence dominates over gas self-gravity during the early clump-forming process. Upon introducing self-gravity in \textbf{Phase III}, the flat density profiles evolve into an isothermal-like profile ($\mathrm{t_4}$, black lines). However, to validate the authenticity of the flat density profile of the turbulence-driven dense cores, self-consistent modeling of gravito-turbulence during gas accretion and halo formation is required, which is beyond the scope of this study.

In $\mathcal{M} = 4$ models, the density and temperature profiles undergo significant changes from \textbf{Phase II} to \textbf{Phase III} simulations; in contrast, clumps in $\mathcal{M} = 8$ models exhibit a similar outer radial profile ($r \gtrsim 0.1$ pc)  between the end of \textbf{Phase II} and \textbf{Phase III} simulations ($\mathrm{t_3}$ and $\mathrm{t_4}$). This suggests that $\mathcal{M} = 4$ turbulence fails to generate a clump-scale structure ($r \ \mathrm{\gtrsim}$ 0.1 pc) without gas self-gravity, but such a structure can be shaped by stronger turbulence ($\mathcal{M} = 8$). Nevertheless, gas self-gravity is necessary to form a dense core with $r \ \mathrm{\lesssim}$ 0.05 pc, where gravity dominates over gas pressure. The $\mathrm{H_2}$ fraction profile changes slightly over time, indicating saturation under turbulent flow without gas self-gravity.

Here, the in-fall rate $R_{in-fall}$ of the clump is defined as 
\begin{equation}
    R_{in-fall}(r) = 4 \pi r^2 {\rho}(r) {v}_{rad}(r),
\end{equation}
where ${\rho}$ is the gas density, and ${v}_{rad}$ is the radial velocity toward the center.
Similar to the density and temperature structure, the prototype of the final configuration of $R_{in-fall}$ has already emerged at the end of \textbf{Phase II} ($\mathrm{t_3}$) only in $\mathcal{M} = 8$. When a GBC is created at the end of the simulations, the in-fall rate exhibits a pattern of increasing with radius but saturating at the outer part, implying the collapse of the clump. Inside the clump, a significant drop in the in-fall rate indicates the formation of an accumulating object at the center. The radius of a dense core is defined as the radius where the in-fall rate starts to decrease substantially within the clump.

\begin{table*}
    \centering
	
    \begin{tabular}{ c|c|c|c|c|c|c|c|c }
    \toprule
    Model & Core &
    ${\rho_{max} \ [g \cdot cm^{-3}]}$ &
    ${\mathbf{x}_{max} \ [pc]}$ &
    $r_{core} \ \mathrm{[pc]}$ &
    $M_{core} \ \mathrm{[M_\odot]}$ &
    $R_{in-fall} \ \mathrm{[M_\odot \cdot yr^{-1}]}$ &
    $L \ \mathrm{[g \cdot cm^2 \cdot s^{-1}]}$ &
    ${f_{kep}}$ 
    \\ 
    
    \midrule
    $M4C3S$ & a &
    2.420 $\mathrm{\times 10^{-16}}$ &
    (0.8,-0.9,-0.2) &
    0.027 &
    80.7 &
    9.589 $\mathrm{\times 10^{-3}}$ &
    3.051 $\mathrm{\times 10^{56}}$ &
    0.159 
    \\
    %%%%%%%%%%%%%%
    
    \midrule
    $M4C4D$ & a & 
    3.773 $\mathrm{\times 10^{-16}}$ &
    (0.3,0.2,-0.5) &
    0.018 &
    52.1 &
    7.414 $\mathrm{\times 10^{-3}}$ &
    3.354 $\mathrm{\times 10^{56}}$ &
    0.413 
    \\
    %%%%%%%%%%%%%%
    
    \midrule
    \multirow{2}*{$M8C3S$} & a &
    3.841 $\mathrm{\times 10^{-16}}$ & 
    (-0.9,-1.2,-0.1) &
    0.019 &
    46.1 &
    3.288 $\mathrm{\times 10^{-3}}$ &
    2.555 $\mathrm{\times 10^{56}}$ &
    0.368 
    \\
    ~ & b & 
    4.333 $\mathrm{\times 10^{-17}}$ &
    (-0.1,0.2,-0.1) &
    0.048 &
    94.5 &
    9.018 $\mathrm{\times 10^{-4}}$ &
    1.402 $\mathrm{\times 10^{57}}$ &
    0.433 
    \\
    %%%%%%%%%%%%%%
    
    \midrule
    \multirow{4}*{$M8C4S$} & a &
    5.077 $\mathrm{\times 10^{-16}}$ & 
    (0.8,0.6,-1.4) &
    0.035 &
    112.2 &
    6.874 $\mathrm{\times 10^{-3}}$ &
    2.768 $\mathrm{\times 10^{57}}$ &
    0.773 
    \\
    ~ & b &
    1.458 $\mathrm{\times 10^{-16}}$ & 
    (0.6,-0.9,-0.3) &
    0.027 &
    48.7 &
    6.454 $\mathrm{\times 10^{-4}}$ &
    4.821 $\mathrm{\times 10^{56}}$ &
    0.536 
    \\
    ~ & c &
    1.252 $\mathrm{\times 10^{-16}}$ &
    (0.5,0.4,-1.2) &
    0.058 &
    80.1 &
    5.565 $\mathrm{\times 10^{-3}}$ &
    5.126 $\mathrm{\times 10^{57}}$ &
    1.844 
    \\
    ~ & d & 
    7.347 $\mathrm{\times 10^{-17}}$ &
    (-1.4,-1.0,0.9) &
    0.038 &
    40.1 &
    7.180 $\mathrm{\times 10^{-4}}$ &
    1.835 $\mathrm{\times 10^{56}}$ &
    0.230 
    \\
    %%%%%%%%%%%%%%
    
    \midrule
    \multirow{4}*{$M8C4D$} & a &
    5.862 $\mathrm{\times 10^{-16}}$ & 
    (-0.2,-0.5,0.8) &
    0.036 &
    174.9 &
    2.963 $\mathrm{\times 10^{-2}}$ &
    2.226 $\mathrm{\times 10^{57}}$ &
    0.315 
    \\ 
    ~ & b & 
    5.836 $\mathrm{\times 10^{-16}}$ &
    (0.4,-0.8,-0.2) &
    0.022 &
    102.1 &
    2.391 $\mathrm{\times 10^{-2}}$ &
    9.523 $\mathrm{\times 10^{56}}$ &
    0.387 
    \\
    ~ & c &
    5.717 $\mathrm{\times 10^{-17}}$ & 
    (-0.1,0.8,-0.4) &
    0.036 &
    46.6 &
    4.432 $\mathrm{\times 10^{-3}}$ &
    6.448 $\mathrm{\times 10^{56}}$ &
    0.663 
    \\
    ~ & d &
    2.082 $\mathrm{\times 10^{-17}}$ & 
    (0.7,-1.4,1.0) &
    0.032 &
    22.7 &
    9.108 $\mathrm{\times 10^{-4}}$ &
    1.144 $\mathrm{\times 10^{56}}$ &
    0.368 
    \\

    \bottomrule
    \end{tabular}

    \caption{Properties of the dense cores.  From left to right, the information includes the model name, core label, maximum density inside the core ${\rho_{max}}$, position of the maximum density ${\mathbf{x}{max}}$ relative to the box center, core radius $r{core}$, gas mass within the core radius $M_{core}$, in-fall rate at the core radius $R_{in-fall}$, core's total angular momentum $L$, and the ratio of rotation to Keplerian velocity $f_{kep}$.}
    \label{tab:DenseCoreProperties}
    
\end{table*}

%%%%%%%%%%%%%%%%%%%%%%%%%%%%%
\subsection{Dense Cores}\label{subsec:DenseCores}

Based on the in-fall rate curve at the end of the simulations, we identify the dense cores of GBCs and list their properties in Table \ref{tab:DenseCoreProperties}. We discuss the possible stellar mass formed within these cores in Section \ref{subsec:StellarMass}. Among the cores with the densest center in each case, the fact that cores in higher $\mathcal{M}$ and $\mathcal{C}$ turbulence have higher maximum density indicates that ${\rho_{\text{max}}}$ depends on the turbulence compressibility.
In our models, core masses range from 22.7 to 174.9 $M_{\odot}$. Higher $\mathcal{M}$ and $\mathcal{C}$ turbulence generate more cores with various masses, naturally creating both low and high-mass cores due to different scales of convergent flows.

By assuming the cores as rigid bodies, we can estimate the rotational velocities are equal to 2 $\mathrm{km \ s^{-1}}$ roughly. For most cores, the ratio of the rotational velocity to its Kepler velocity ${f_{\text{kep}}}$ is lower than 0.8, except for $M8C4S$-$c$. Generally, ${f_{\text{kep}} < 1}$ suggests that the rotating structure can be bound by gravity. The core of $M8C4S$-$c$ has ${f_{\text{kep}} = 1.844}$, implying that the centrifugal force will break this core and lead to fragmentation.

%%%%%%%%%%%%%%%%%%%%%%%%%%%%%%%%%%%%%%%%%%%%%%%%%%%%%%%%%%%%%%
%%%%%%%%%%%%%%%%%%%%%%%%%%%%%%%%%%%%%%%%%%%%%%%%%%%%%%%%%%%%%%

\section{Discussions} \label{sec:Discussions}

%%%%%%%%%%%%%%%%%%%%%%%%%%%%%
\subsection{Summary of the Turbulent Primordial Cloud Model}\label{subsec:SummaryOfTheTurbulentPrimordialCloudModel}

In the \textbf{Phase I} simulation, a uniform primordial cloud is stirred by the stochastic forcing field. As turbulent structures gradually develop, $\mathrm{H_2}$ fraction grows to $\mathrm{\sim \, 10^{-4}}$, and the gas temperature reaches its maximum value. Subsequently, the $\mathrm{H_2}$ fraction continues to increase to $\mathrm{\sim \, 10^{-3}}$, and the temperature attains a minimum equilibrium through molecular hydrogen cooling. Most of the gas has a Mach number distributed around the model's corresponding $\mathcal{M}$. Once turbulence fully develops, a DM halo potential is introduced, and the driven force of turbulence is weakened as the simulation enters \textbf{Phase II}. Eventually, the virialization parameter of the gas system drops to $\mathrm{\sim \, 0}$ at the end of \textbf{Phase II}. The gas distribution evolves from an isotropic configuration to a centrally contracted structure due to the gravitational potential well of the halo. After removing the stochastic turbulence at the beginning of \textbf{Phase III}, gas self-gravity becomes the dominating force in determining the gas dynamics on scales $<$ 0.1 pc. The small-scale gas accretion due to self-gravity not only collapses the existing GBCs to form dense cores at the center but also condenses relatively less compact gas clusters, leading to their growth and the generation of dense cores. Additionally, the rotation of these dense cores could influence subsequent Pop~III SF \citep{ekstrom2008effects, yoon2012evolution, stacy2013rotation}.

Weak turbulence ($\mathcal{M} \leq 2$) or less compressive turbulence ($\mathcal{C} \leq 2$) cannot produce any GBCs after removing the stochastic forcing turbulence. Turbulence with $\mathcal{M} \geq 4$ and $\mathcal{C} \geq 3$ forms GBCs that grow further into dense cores at the end of the simulation. However, intermediate turbulence with $\mathcal{M} = 4$ or $\mathcal{C} = 3$ generates either one clump or no clump due to the random nature of stochastic forcing turbulence. The $\mathcal{M} = 4$ models that have formed clumps create only a single dense core in the primordial cloud. Meanwhile, $\mathcal{M} = 8$ can form multiple dense cores with various core masses, and higher $\mathcal{C}$ can increase the number of clumps.

%%%%%%%%%%%%%%%%%%%%%%%%%%%%%
\subsection{From Dense Cores to Stellar Masses}\label{subsec:StellarMass}

The ultimate goal of our simulation is to determine the mass of Pop~III stars. However, within the limitations of our simulation, the smallest high-density objects we can achieve are the dense cores derived from the in-fall rate. 
These cores can not be properly evolved further in our current simulations due to the lack of small-scale ($\rho \, \mathrm{ > 10^{-15} \, g \, cm^{-3}}$) physics and finer spatial resolution. 
Instead, we infer the possible stellar mass by assuming the mass function of the stars resembles the core mass function (CMF) based on the result of \citet{guszejnov2015mapping}.
Consequently, the profile of CMF and initial mass function (IMF) are similar except for a $\sim$ 1/3 shift of the peak value. 
Therefore, we divide the mass of cores by a factor of three to obtain the expected stellar mass range of $8 -59 \mathrm{M_{\odot}}$. 
This mass range agrees with the typical Pop~III stellar mass inferred from the EMP stars observation \citep{susa2014mass,ishigaki2018initial}.

On the other hand, different from the EMP star observation, \citet{abe2021formation,chen2022impact} suggests another method to probe the Pop~III IMF through the first supernovae and galaxies, potentially observable targets to the James Webb Space Telescope (JWST). 

%In sum, the forthcoming JWST observations of the early universe will shed light on the formation of Pop~III stars.

%%%%%%%%%%%%%%%%%%%%%%%%%%%%%
\subsection{Formation of Very Massive Pop~III stars}\label{subsec:Observation}

In our $\mathcal{M} = 2$ or low $\mathcal{C} = 2$ models, they can not produce adequate high-density regions on the small scale, but the turbulent motion prevents the direct collapse of the cloud. Accordingly, weak or less compressive turbulence fails to fragment the cloud into smaller dense substructures.
Nevertheless, as the cooling mechanism releases turbulence energy, the primordial cloud may eventually undergo a large-scale collapse similar to the previous cosmological simulations, resulting in more massive Pop~III stars over $100 M_{\odot}$.
Therefore, turbulence nature can alter the typical stellar mass in a mini-halo, bridging the low-mass and high-mass end of Pop~III stars.

\subsection{Possible Effects of Magnetic fields}\label{subsec:MagneticField}
In this study, we have not considered the effects of magnetic fields, which can be amplified through the small-scale dynamo in turbulent primordial clouds.  Magnetic fields play a crucial role in transferring the angular momentum of the proto-stellar disk, enabling a proto-star to accrete gas successively, and they are significant contributors to the SF process \citep{crutcher12, krum19}.

Recent studies \citep{mckee2020magnetic, sharda2020importance, stacy2022magnetic, Saad22} have investigated the impact of magnetic fields on Pop~III SF using high-resolution magnetohydrodynamic (MHD) simulations. Their results suggest that magnetic fields suppress the formation of low-mass Pop~III stars by delaying gas collapse and inhibiting fragmentation within star-forming disks. However, numerical viscosity and resistivity in these simulations are still orders of magnitude larger than the physical values. Additionally, none of these studies consider non-ideal MHD effects, such as ambipolar diffusion, which can alter the resistivity of the fluid and influence small-scale dynamo processes. Consequently, the evolution of magnetic fields within the Pop~III star-forming cloud and their impact on Pop~III SF remains unclear.

%%%%%%%%%%%%%%%%%%%%%%%%%%%%%
\subsection{Further Improvements for the Current Model}\label{subsec:PossibleImprovement}

So far, we have explored the impact of different turbulence on the primordial cloud through artificially driven turbulence. However, the turbulence structure is not self-consistently generated in our simulations. Therefore, simulating gravitational-driven turbulence during the mini-halo formation may provide insights into the nature of turbulence in the context of Pop~III SF.

Our simulation stops at $\rho_{\text{max}} \, \sim 10^{-16} \mathrm{g \, cm^{-3}}$ since we have not included all the relevant small-scale physics and finer spatial resolution. If the simulation continued evolving the dense core, a fully molecular core would rapidly form via three-body reactions and become optically thick. This would eventually lead to the formation of one or multiple proto-stellar cores. The gas in the cores would then accrete onto proto-stars through proto-stellar disks, and radiative feedback would determine the subsequent accretion process. Additionally, a comprehensive understanding of the formation of Pop III stars and their mass distribution necessitates the consideration of the influence of magnetic fields.
Therefore, in our future models, we plan to incorporate relevant physics such as a disk model, radiative feedback, magnetic fields, and SF to obtain the Pop~III stellar mass self-consistently.

%%%%%%%%%%%%%%%%%%%%%%%%%%%%%%%%%%%%%%%%%%%%%%%%%%%%%%%%%%%%%%
%%%%%%%%%%%%%%%%%%%%%%%%%%%%%%%%%%%%%%%%%%%%%%%%%%%%%%%%%%%%%%

\section{Conclusion} \label{sec:Conclusion}

In this study, we have introduced a numerical method to replicate the turbulent structure within primordial clouds by combining an external stochastic forcing field and primordial gas composition. Our investigation delves into understanding how turbulence influences the gas structure and potential star-forming sites under different turbulence parameters. The simulations reveal that only sufficiently strong ($\mathcal{M} \geq 4$) and highly compressive ($\mathcal{C} \geq 3$) turbulence can generate locally fragmentary structures ($\mathrm{\gtrsim}$ 0.1 pc) within the primordial cloud. Moreover, turbulence with stronger compressibility (higher $\mathcal{M}$ and $\mathcal{C}$) forms more gravitationally bound gas clumps, ultimately leading to dense cores with masses ranging from 22.7 $\mathrm{M_{\odot}}$ to 174.9 $\mathrm{M_{\odot}}$ at the end of our simulations.

Considering the relationship between the Core Mass Function (CMF) and Initial Mass Function (IMF) discovered by \citet{guszejnov2015mapping}, the expected final stellar mass range of these dense cores aligns with $\sim 8-59 \, \mathrm{M_{\odot}}$ roughly, which agrees with observations of Extremely Metal-Poor (EMP) stars. This result instills confidence in our future work on forming less massive Pop~III stars in turbulent primordial clouds, aiming to reconcile the existing discrepancy between simulations and observations.

For the $\mathcal{M} = 2$ or $\mathcal{C} = 2$ turbulence, the scenario is similar to the monolithic collapse of primordial gas in the mini-halo suggested by the previous cosmological simulations.
Therefore, our turbulence model can bridge between the low and high mass Pop~III star formation.
In the subsequent work, our model will be improved with more realistic initial conditions and microphysics.
By self-consistently simulating turbulent structures of primordial clouds and following star formation, we will probe the characteristic mass and IMF of Pop~III stars.
With the JWST observation and sophisticated models, we will soon peak into the cosmic dawn by understanding the birth of the first stars.

%%%%%%%%%%%%%%%%%%%%%%%%%%%%%%%%%%%%%%%%%%%%%%%%%%%%%%%%%%%%%%
%%%%%%%%%%%%%%%%%%%%%%%%%%%%%%%%%%%%%%%%%%%%%%%%%%%%%%%%%%%%%%

\section*{Acknowledgements}

Tang, Ching-Yao acknowledges his colleague Mr. Lin, Sanctity and Prof. Lee, Yueh-Ning for their support. This research is supported by the National Science and Technology Council under grant no. MOST 110-2112-M-001-068-MY3 and the Academia Sinica, Taiwan under a career development award under grant no. AS-CDA-111-M04. Our computing resources were supported by the National Energy Research Scientific Computing Center (NERSC), a U.S. Department of Energy Office of Science User Facility operated under Contract No. DE-AC02-05CH11231,  and the TIARA Cluster at the Academia Sinica Institute of Astronomy and Astrophysics (ASIAA).

%%%%%%%%%%%%%%%%%%%%%%%%%%%%%%%%%%%%%%%%%%%%%%%%%%
\section*{Data Availability}

The simulation data underlying this article were generated from NERSC supercomputer Cori. Due to the huge size of data volume,  the derived data generated in this research will be shared on reasonable request to the corresponding author.



%\bibliographystyle{mnras}
%\bibliography{refs} % if your bibtex file is called example.bib

%\begin{thebibliography}{99}
%\end{thebibliography}

%%%%%%%%%%%%%%%%%%%%%%%%%%%%%%%%%%%%%%%%%%%%%%%%%%

%%%%%%%%%%%%%%%%% APPENDICES %%%%%%%%%%%%%%%%%%%%%

%\appendix


%If you want to present additional material which would interrupt the flow of the main paper, it can be placed in an Appendix which appears after the list of references.

%%%%%%%%%%%%%%%%%%%%%%%%%%%%%%%%%%%%%%%%%%%%%%%%%%

% Don't change these lines
\bsp	% typesetting comment
\label{lastpage}
\end{document}